\documentclass[12pt,oneside, a4paper]{article}

\ifx\pdfoutput\undefined
\usepackage[dvips,bookmarks=false]{hyperref}	
\else
\usepackage{hyperref}	
\fi
\hypersetup{colorlinks,bookmarksopen,bookmarksnumbered,citecolor=blue,
linkcolor=black,pdfstartview=FitH,urlcolor=blue}

\usepackage{graphicx}
\usepackage{amssymb}
\usepackage{cite}
\usepackage{bm}
\usepackage{indentfirst}
\usepackage{amsmath}
\usepackage{hhline}
\usepackage{multirow}
\usepackage{slashed}

\allowdisplaybreaks

\renewcommand{\cal}{\mathcal}

\renewcommand{\bar}{\overline}

\usepackage{ulem}

\begin{document}

\begin{titlepage}

\begin{flushright}
IPMU21-0032\\
KANAZAWA-21-07
\end{flushright}

\begin{center}

\vspace{1cm}
{\large\textbf{
Vanishing or non-vanishing rainbow?\\ 
Reduction formulas of electric dipole moment
}
}
\vspace{1cm}

\renewcommand{\thefootnote}{\fnsymbol{footnote}}
Motoko Fujiwara$^{1}$\footnote[1]{motoko@eken.phys.nagoya-u.ac.jp}
,
Junji Hisano$^{1,2,3}$\footnote[2]{hisano@eken.phys.nagoya-u.ac.jp}
,
Takashi Toma$^{4,5}$\footnote[3]{toma@staff.kanazawa-u.ac.jp}

\vspace{5mm}

\textit{
$^1${Department of Physics, Nagoya University,\\ Furo-cho Chikusa-ku, Nagoya, 464-8602 Japan}\\
$^2${Kobayashi-Maskawa Institute for the Origin of Particles and the Universe,\\ Nagoya University,Furo-cho Chikusa-ku, Nagoya, 464-8602 Japan}\\
$^3${Kavli IPMU (WPI), UTIAS, University of Tokyo, Kashiwa, 277-8584, Japan}\\
$^4${Institute of Liberal Arts and Science, Kanazawa University,\\ Kakuma-machi, Kanazawa, 920-1192 Japan}\\
$^5${Institute for Theoretical Physics, Kanazawa University, Kanazawa, 920-1192 Japan}\\
}

\vspace{8mm}

\abstract{
In this paper, we derive a simplified formula of electric dipole moments (EDMs) of a fermion. 
In the Standard Model, it is well-known that non-trivial cancellations 
between some rainbow-type diagrams induced by $W$ boson exchanges occur in the calculation of the neutron
EDM at the two-loop level due to the gauge symmetry. 
The fermion self-energy and the vertex correction are related through the Ward-Takahashi identity, 
and this relation causes the exact cancellation of the EDM. 
We derive EDM formulas for a more general setup  
by introducing the form factors for the fermion self-energy and the vertex correction so that the derived formulas can be applicable to a larger class of models.
We conclude that the non-zero EDM contributions are induced from rainbow-type diagrams with the chirality flipping effects for internal fermions. 
We also discuss the other possible generalization of the EDM calculation which is applicable to the other classes of models.
}

\end{center}
\end{titlepage}

\renewcommand{\thefootnote}{\arabic{footnote}}
\newcommand{\bhline}[1]{\noalign{\hrule height #1}}
\newcommand{\bvline}[1]{\vrule width #1}

\setcounter{footnote}{0}

\setcounter{page}{1}


\section{Introduction}
\label{sec:1}

The CP violation is the key to understand the nature of the universe dominated by baryons.
In the Standard Model (SM), the possible CP violating source is the Cabbibo-Kobayashi-Maskawa (CKM) matrix
that appears in the processes mediated by $W$ bosons between up-type and down-type quarks.
Such CP violating processes are suppressed by the small Jarlskog invariant in the electroweak theory~\cite{Jarlskog:1985ht, Jarlskog:1985cw}. 

\begin{figure}[htb]
\centering
\includegraphics[width=8cm]{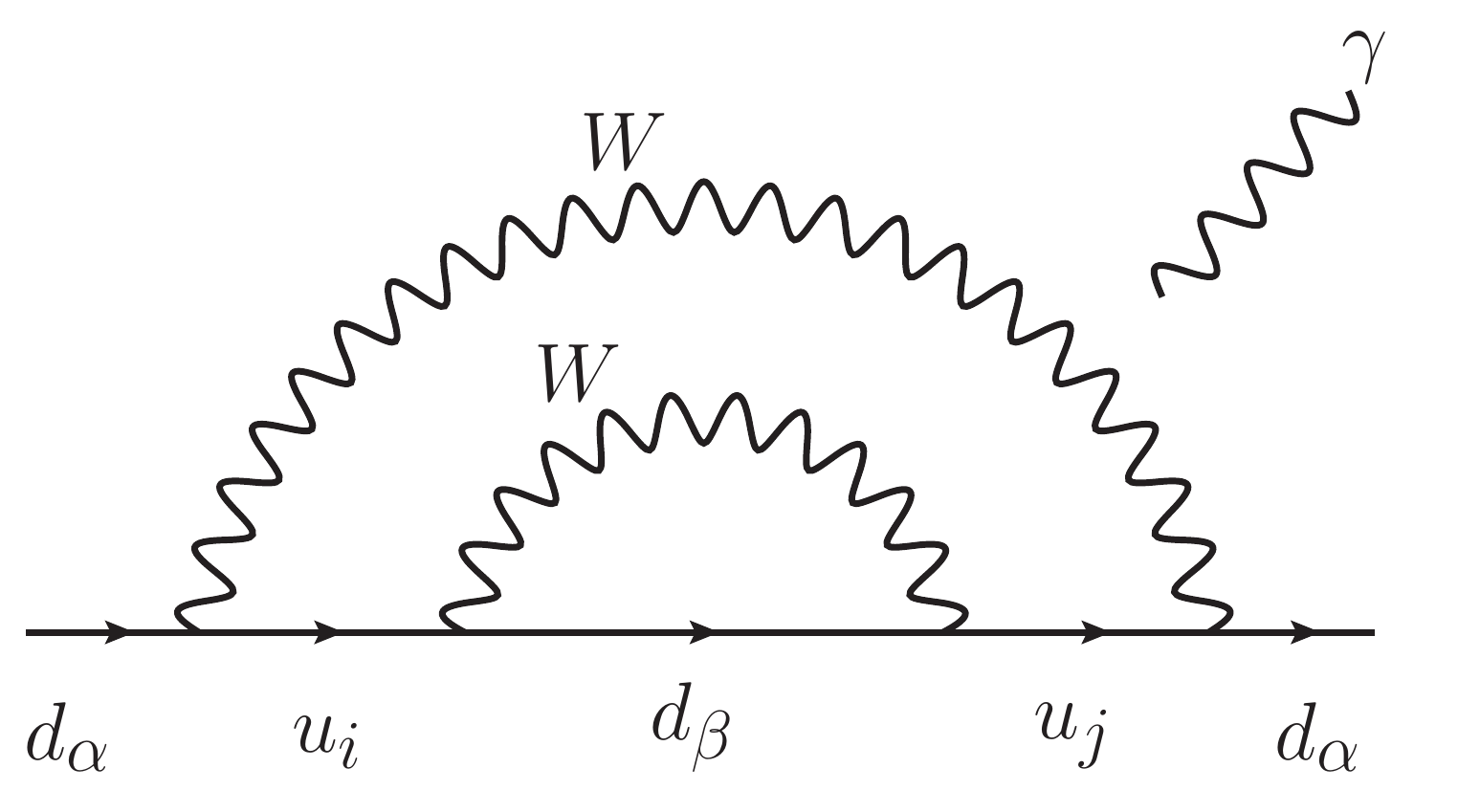}
\caption{Two-loop diagram with $W$ boson propagation in the SM.
The on-shell photon can couple to arbitrary charged particles.}
\label{fig:W-loop_intro}
\end{figure}

The electric dipole moment (EDM) is one of the observables connected with CP violation, and is being explored by various experiments. 
The current upper bound on the neutron EDM is given by the nEDM Collaboration while the bound on the electron EDM is given by the ACME Collaboration, which are 
$|d_n|\leq1.8\times10^{-26}~e\hspace{0.1cm}\mathrm{cm}$~\cite{Abel:2020gbr} and $|d_e|\leq1.1\times10^{-29}~e\hspace{0.1cm}\mathrm{cm}$~\cite{Andreev:2018ayy} at $90\%$ confidence level, respectively. 
For future prospects, the TUCAN EDM experiment aims to measure the neutron EDM at $10^{-27}~e\hspace{0.1cm}\mathrm{cm}$~\cite{Martin:2020lbx} and 
the ACME Collaboration will update the bound for electron EDM at $\mathcal{O}(10^{-30})~e\hspace{0.1cm}\mathrm{cm}$~\cite{Kara:2012ay, edm_future}. 
Theoretically, the EDM is induced from the flavor diagonal effective Hamiltonian induced by quantum corrections.
For the neutron EDM in the SM,
a leading order diagram is naively expected to be the two-loop level that is proportional to
the fourth order of the CKM matrix elements.
Figure~\ref{fig:W-loop_intro} shows the two-loop ``rainbow-type'' diagram, 
a diagram with a irreducible structure in the outer loop, in the SM where the photon couples to arbitrary charged particles.

However, 
this naive expectation is failed as shown by E. P. Shabalin~\cite{Shabalin:1978rs}.
The author has performed the explicit calculations in the SM
and shown that the neutron EDM induced from these two-loop diagrams totally cancels out.
Therefore, the leading order contribution to the neutron EDM is induced by the three-loop level.\footnote{It is rather induced from the bound state effect
that is estimated to be $|d_n|\sim 10^{-31}~e\hspace{0.1cm}\mathrm{cm}$~\cite{Dar:2000tn, Mannel:2012qk}.}
For the charged leptons in the SM, the EDMs are further suppressed and induced at four-loop order because the CP violation has to be brought 
from the quark sector~\cite{Pospelov:2005pr, Fukuyama:2012np}.
In addition to the SM discussion, 
the similar cancellation of charged lepton EDMs in Type-I seesaw models has also been discussed in Refs.~\cite{Ng:1995cs, Archambault:2004td, Chang:2004pba}.

According to Ref.~\cite{Shabalin:1978rs},
this non-trivial EDM cancellation between the diagrams shown in Fig.~\ref{fig:W-loop_intro} can be attributed to the electromagnetic gauge symmetry.
These rainbow-type diagrams
have the structures of one-loop sub-diagrams that induce the fermion self-energy and the vertex correction.
These contributions are related to each other
through the Ward-Takahashi identity of the gauge symmetry.
This is one of the most essential points to understand the neutron EDM cancellation at the  two-loop level in the SM. 
We also expect the same technique using the Ward-Takahashi identity should be useful to derive the EDM formulas not only in the SM but also in a  wider class of models. 
%
%
%
%

In this paper, we generalize Shabalin's explicit calculation which can be applied to rainbow-type diagrams with a more general setup.
Namely,
we introduce the effective off-shell couplings for fermions
that are read out from the self-energy and the vertex correction.
These terms have non-trivial relations through the Ward-Takahashi identity.
We also introduce the form factors to express the fermion self-energy and the vertex correction,
and explicitly show which terms induce non-zero EDM contributions.
With this general setup, 
we finally derive reduced formulas of the EDM 
that can be applicable to a larger class of models including the SM. 
The exact cancellation of the EDM between rainbow-type diagrams in the SM is recovered from the derived formulas. 
Furthermore, we discuss the other possible generalization of the EDM calculation for the models with multi-scalars. 

The remaining parts of the paper are organized as follows.
In Sec.~\ref{sec:2}, we derive the reduction formulas for EDMs induced by two-loop or higher order diagrams that include sub-diagram structures. 
In Sec.~\ref{sec:3}, we apply our formulas of the EDM to some specific models 
such as the SM, the scotogenic model and the singlet-triplet extended model. 
Our reduction formulas are general enough to not only reproduce the EDM cancellation between the rainbow-type diagrams in the SM but also provide the predicted EDMs in a larger class of models beyond the SM.
In Sec.~\ref{sec:4}, we also consider the other possible generalizations of the EDM calculation for the models with multi-scalars. 
Our conclusions are given in Sec.~\ref{sec:conclusion}.

\section{Reduction formulas of EDM}
\label{sec:2}

In this section, 
we derive reduction formulas for EDM that are induced from the diagrams with sub-diagram structures. 
First, 
we briefly review the SM calculation of the neutron EDM diagrams at the two-loop level as discussed in Ref.~\cite{Shabalin:1978rs}. 
Second, 
we summarize the essential points in the SM calculation 
and derive the EDM reduction formulas with a more general setup.

\subsection{EDM cancellation in the Standard Model}
We focus on the electroweak interactions between up-type and down-type quarks 
as the CP violation sources in the SM, and calculate the EDM based on the following Lagrangian
\begin{align}
  {\cal  L}
  &=  
 -  \Bigg(\frac{g_2}{\sqrt{2}}  V_{i  \alpha} W_{\mu}^{+}  \bar{u}_i  \gamma^{\mu} P_L  d_\alpha + \mathrm{h.c.}\Bigg) 
 -e Q_u  A_\mu  \bar{u}_i  \gamma^\mu  u_i,
  ~~~~( i,  \alpha  =  1, 2, 3 ),
\label{eq:lag_sm}
\end{align}
where 
$Q_u$ is the electric charge for the up-type quarks, 
$g_2$ is the SU(2)$_L$ gauge coupling 
and $V_{i  \alpha}$ denotes the CKM matrix elements. 
For the later convenience, 
we express the generation indices as roman characters ($i,  j$) for the up-type quarks, 
and greek characters ($\alpha,  \beta$) for the down-type quarks, respectively. 
In the EDM calculation, it suffices to focus on the CP violating terms that are up to the first order of photon momentum as pointed out in Ref.~\cite{Shabalin:1978rs}. 
One can naively expect that the leading order CP violating terms arise from the amplitude at ${\cal  O}  ( \alpha_2^2 )$ $( \alpha_2  =  g_2^2/ 4  \pi )$ 
as shown in Fig.~\ref{fig:W-loop_intro}.\footnote{The discussion for the diagrams with the  external up-type quarks is parallel. 
}
The on-shell photon can couple to arbitrary charged particles in the diagram. 
These amplitudes are proportional to the four CKM matrix elements, 
and the non-zero CP phases arise from their imaginary parts, 
\begin{align}
  \mathrm{Im}  \left[ V^*_{j  \alpha}  V_{j  \beta}  V^*_{i  \beta}  V_{i  \alpha} \right], 
\end{align}
which is the Jarlskog invariant~\cite{Jarlskog:1985ht, Jarlskog:1985cw}.
If we exchange the internal fermion flavor indices $i  \leftrightarrow  j$, 
this Jarlskog invariant acquires $(-1)$ factor as shown below.
\begin{align}
  \mathrm{Im}  \left[ V^*_{j  \alpha}  V_{j  \beta}  V^*_{i  \beta}  V_{i  \alpha} \right]
  &\xrightarrow[]{i  \leftrightarrow  j}
  -  \mathrm{Im}  \left[ V^*_{j  \alpha}  V_{j  \beta}  V^*_{i  \beta}  V_{i  \alpha} \right]. 
\end{align}
Thanks to this anti-symmetricity of the Jarlskog invariant, 
we can further narrow down the diagrams that may induce non-zero EDM contributions. 
If the photon couples to the outer $W$ boson loop, 
the expression of the amplitude is anti-symmetric under $i  \leftrightarrow  j$. 
Therefore, this contribution cancels out after summing up all the flavor indices. 

\begin{figure}[tb]
\centering

\begin{minipage}{1\hsize}
\centering
\begin{minipage}{0.45\hsize}
\centering
\includegraphics[width=6cm]{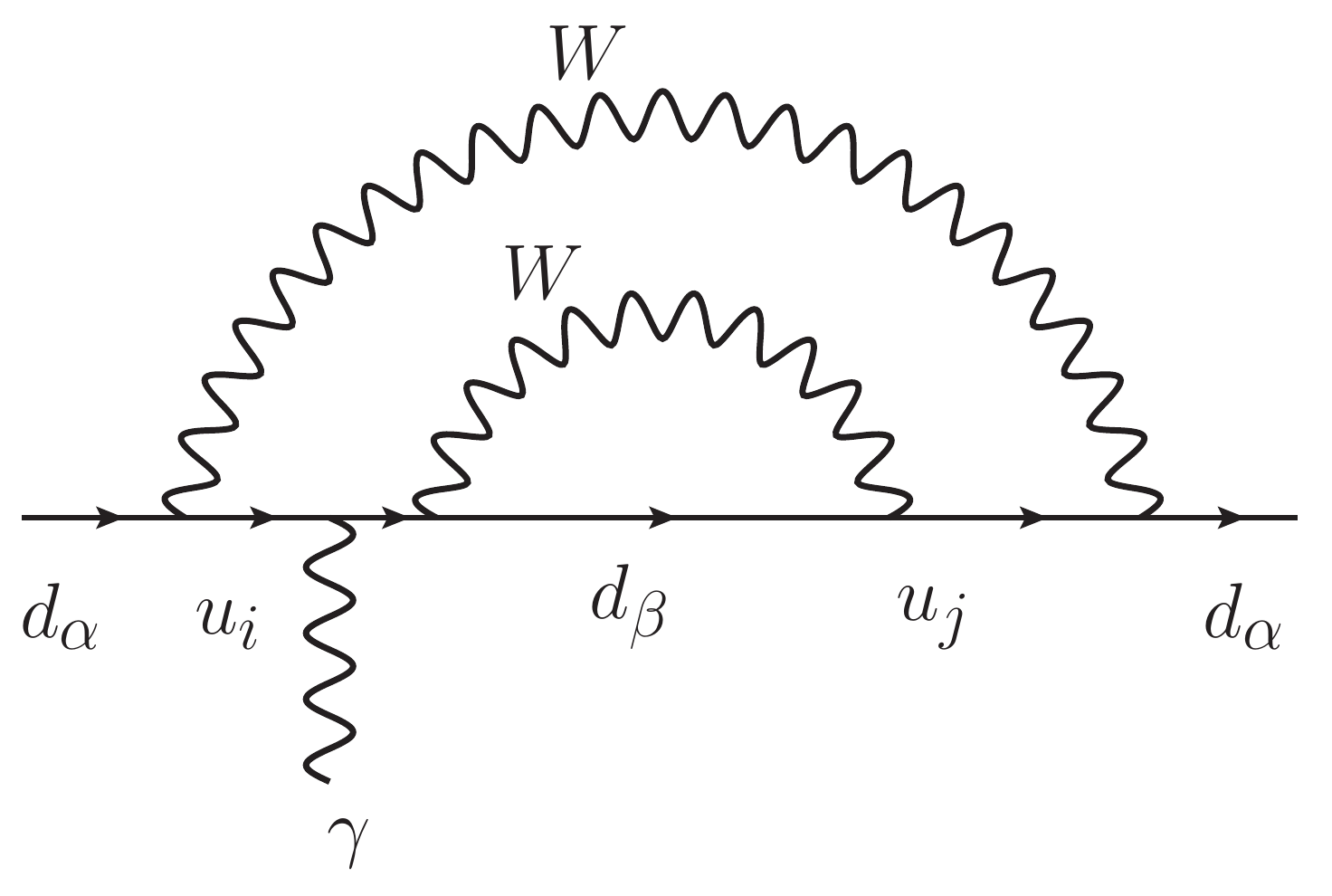}
\\
~\\
\end{minipage}
\begin{minipage}{0.45\hsize}
\centering
\includegraphics[width=6cm]{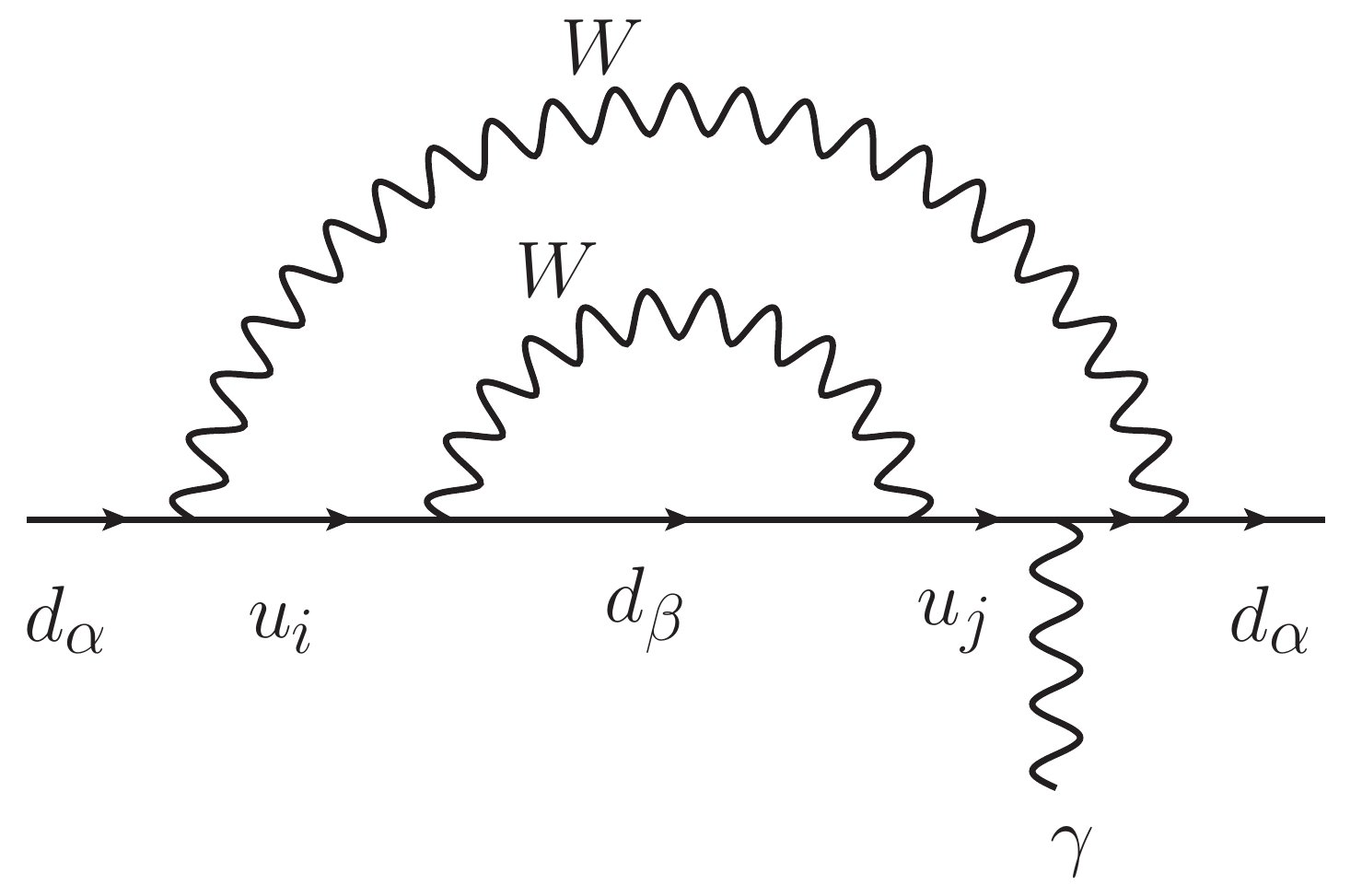}
\\
~\\
\end{minipage}
\end{minipage}

\begin{minipage}{1\hsize}
\centering
\begin{minipage}{0.45\hsize}
\centering
\includegraphics[width=6cm]{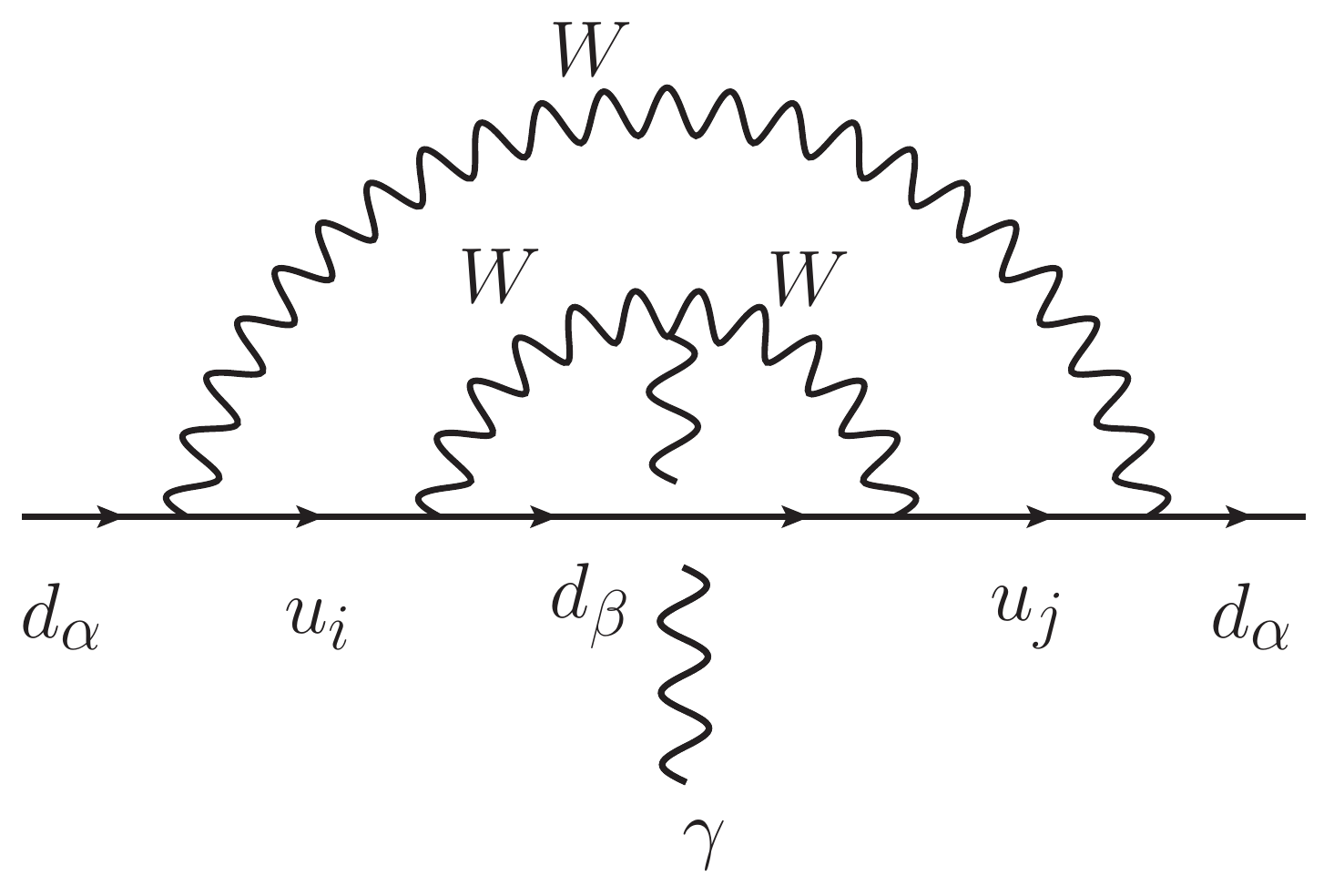}
\\
\end{minipage}
\begin{minipage}{0.45\hsize}
\centering
\includegraphics[width=6cm]{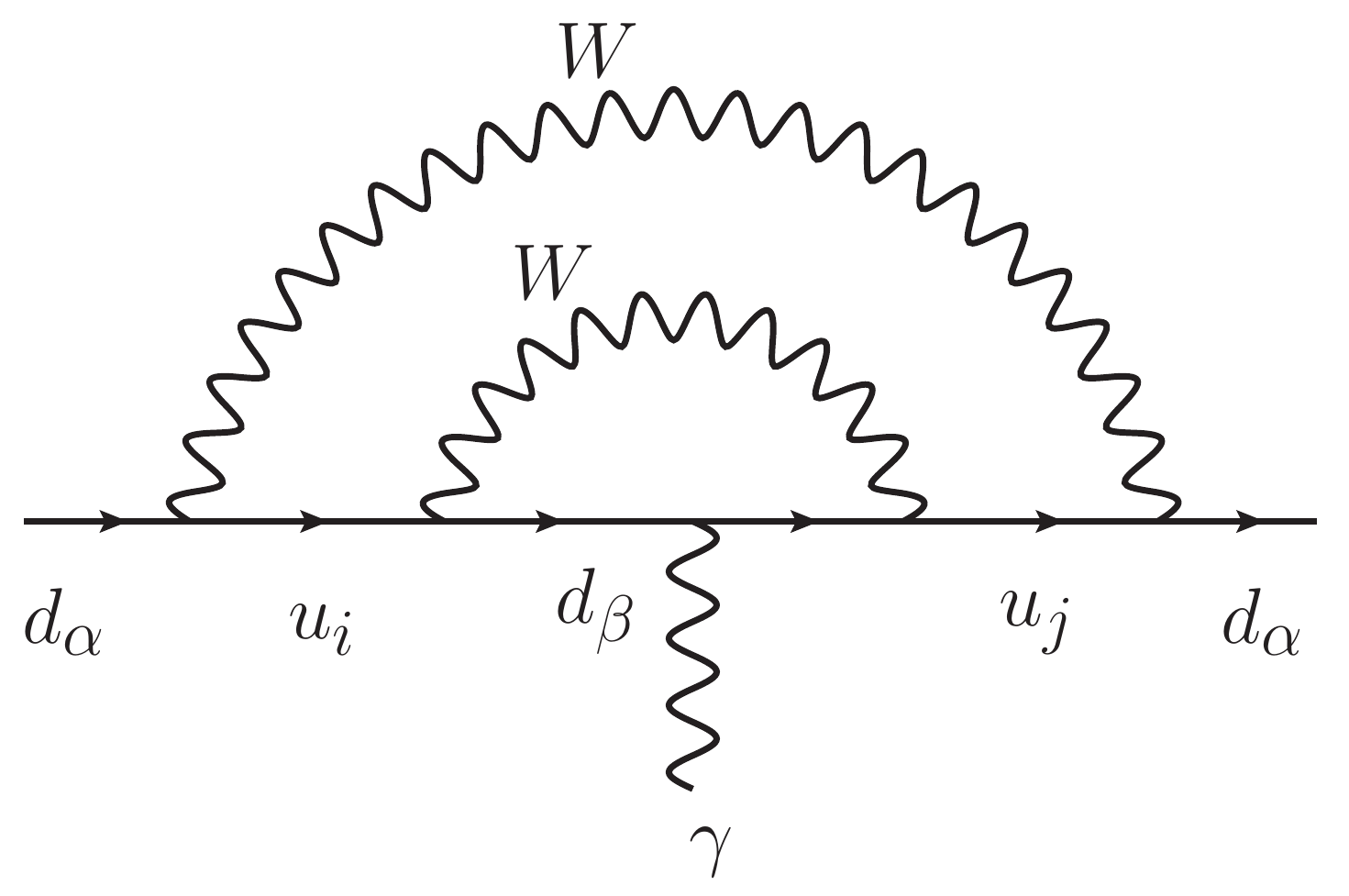}
\\
\end{minipage}
\end{minipage}

\caption{Variations of diagrams that have a chance to induce non-zero EDM. 
We find the self-energy structure for $u_i$ and $u_j$ from the diagrams in the upper row.
On the other hand, we find vertex correction from the diagrams in the lower row.
}
\label{fig:W-loop_w_photon}
\end{figure}

The remaining diagrams are shown in Fig.~\ref{fig:W-loop_w_photon}. 
We need to have $i  \ne  j$ in these diagrams to pick up non-zero CP phases. 
There are four variations of the diagrams that a photon can attach to the charged particles. 
These individual diagrams induce non-zero EDM contributions. 
However, 
in Ref.~\cite{Shabalin:1978rs}, 
it was pointed out that a non-trivial cancellation occurs between these diagrams. 
The point is 
to find the sub-diagram structures of the self-energy and vertex correction for the up-type quarks, $u_i$ and $u_j$.  
These self-energy and vertex correction are related to each other through the Ward-Takahashi  identity of the electromagnetic gauge symmetry. 
This is one of the most crucial points to understand the EDM cancellation in these diagrams. 
At the same time, 
we expect the same reduction technique using the Ward-Takahashi identity 
can be applied for a more general setup other than the SM. 
In the succeeding section, 
we summarize the points in the SM calculation and show how to generalize our calculation setup.

\subsection{Derivation of EDM reduction formulas}
The important points for neutron EDM cancellation at the two-loop level in the SM are summarized below~\cite{Shabalin:1978rs}.
\begin{itemize}
  \item  The amplitude is expanded by the photon momentum up to the first order, and the EDM contributions arise from the first order term of the on-shell photon momentum.
  \item  The overall CKM factors (Jarlskog invariant) have anti-symmetricity under the exchange of the internal fermion flavor indices ($i  \leftrightarrow  j$).
  \item  The Ward-Takahashi identity works between the self-energy and the vertex correction for internal fermions. 
\end{itemize}
Note that the second point in the above is rather model-dependent and can be adapted only for a narrow class of models involving the SM. 
In the following generalization of the EDM calculation, 
it can be reinterpreted as a result of imposing the hermiticity of the self-energy and the vertex correction 
in the effective Lagrangian. 

\begin{table}[tb]
 \renewcommand{\arraystretch}{1.5}
 \centering
 \begin{tabular}{|c||c|c|c|}
  \hline
  &  $f_\alpha$  &  $\psi_i$  &  $\phi$\\\hhline{|=#=|=|=|}
  Spin  &  $1/2$  &  $1/2$  &  $0$\\  \hline
  electric charge  &  $Q_f$  &  $Q_{\psi}$  &  $Q_\phi$\\  \hline
 \end{tabular}
 \caption{Particle contents of the general model we consider where the electric charges $Q_f$, $Q_\psi$ and $Q_\phi$ can be arbitrary values.}
 \label{tab:contents_scalar-loop}
\end{table}

\begin{figure}[t]
\begin{center}
\includegraphics[scale=1.1]{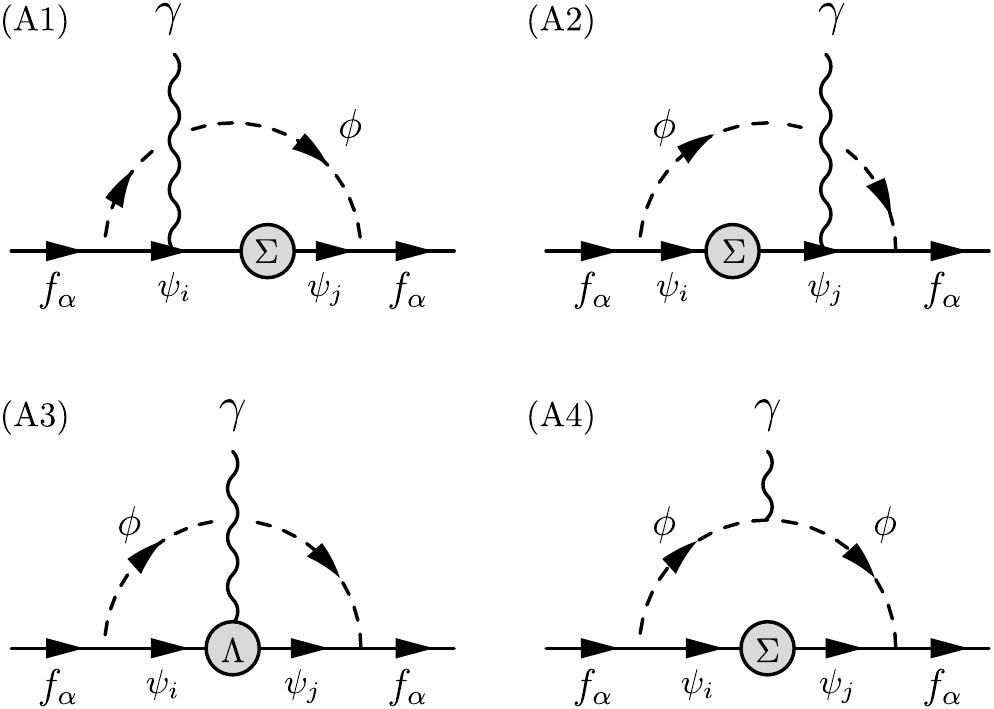}
\caption{Feynman diagrams potentially inducing non-zero EDMs of the fermion $f_\alpha$ with the multi-fermions $\psi_i$ and a scalar $\phi$.}
\label{fig:diagram_fermion_scalar}
\end{center}
\end{figure}

To generalize our calculation setup 
without loss of the essence in the SM calculation, 
we introduce the following effective Lagrangian
\begin{align}
\mathcal{L}_\mathrm{eff}=&-\Big(y_{i\alpha}\phi^*\overline{\psi_i}P_Lf_{\alpha}+\mathrm{H.c.}\Big)
 -\overline{\psi_i}\Sigma_{ij}\psi_j
-eA_{\mu}\overline{\psi_i}\Lambda_{ij}^{\mu}\psi_j\nonumber\\
&-eQ_{\psi}A_{\mu}\overline{\psi_i}\gamma^{\mu}\psi_i
+ieQ_{\phi}A_{\mu}\big(
\phi\partial^{\mu}\raisebox{0.33cm}{\tiny{$\hspace{-0.4cm}\leftrightarrow$}\hspace{0.1cm}}\phi^*
\big),
\label{eq:lag}
\end{align}
where $\phi\partial^{\mu}\raisebox{0.33cm}{\tiny{$\hspace{-0.4cm}\leftrightarrow$}\hspace{0.1cm}}\phi^*=\phi\partial^{\mu}\phi^*-\phi^*\partial^{\mu}\phi$ and 
\begin{align}
\Lambda_{ij}^{\mu}=\sum_{p}Q_p\big(\Lambda_p\big)_{ij}^{\mu}.
\end{align}
The $W$ boson, up-type and down-type quarks in the SM are replaced by a scalar $\phi$, internal fermion $\psi_i$ and external fermion $f_\alpha$, respectively. 
The spin and electric charges of these particles are summarized in Tab.~\ref{tab:contents_scalar-loop}. 
The Yukawa couplings $y_{i\alpha}$ are possible sources of a CP violation. 
The symbols $\Sigma_{ij}$ and $\Lambda_{ij}^{\mu}$ are the self-energy and the vertex correction induced at the one-loop or higher loop level. 
The subscript $p$ in the vertex correction denotes the particle attaching a photon in the loop. 
All the possible vertex corrections are summed up with the electric charge of the particle $p$ ($Q_p$). 
From this effective Lagrangian, one can draw the diagrams which potentially induce non-zero EDMs of the fermion $f_\alpha$ as shown in Fig.~\ref{fig:diagram_fermion_scalar}.\footnote{In Fig.~\ref{fig:diagram_fermion_scalar}, we also show the diagram with photon that couples to the outer scalar loop, which corresponds to the vector loop diagram not to be shown in Fig.~\ref{fig:W-loop_w_photon}. We explicitly check the cancellation of EDM from this diagram in the view from the hermiticity of the self-energy and the vertex correction. }

One can also consider the diagrams with a vector boson $X$ in the outer loop instead of the scalar $\phi$. However the calculation is essentially the same 
with the scalar case, and the derived formula for the vector boson case will be rather redundant. 
Thus we will give our calculation for the scalar loop below based on the Lagrangian in Eq.~(\ref{eq:lag}), 
and we provide only the final result for the vector boson in Sec.~\ref{sec:vector_loop}, which is obtained by merely replacing the corresponding factors in the scalar case. 
Although we have considered the left chirality of the fermion $f_\alpha$ in Eq.~(\ref{eq:lag}), one can consider right-chirality instead. 
In this case, the following calculations simply change with opposite chirality. 
Note that if both chiralities are involved at tree-level, non-zero EDM of the fermions $f_\alpha$ are induced at one-loop levels, thus our reduction of the EDM calculation which will be shown 
in the following is spoiled. 

\begin{figure}[t]
\begin{center}
\includegraphics[scale=0.4]{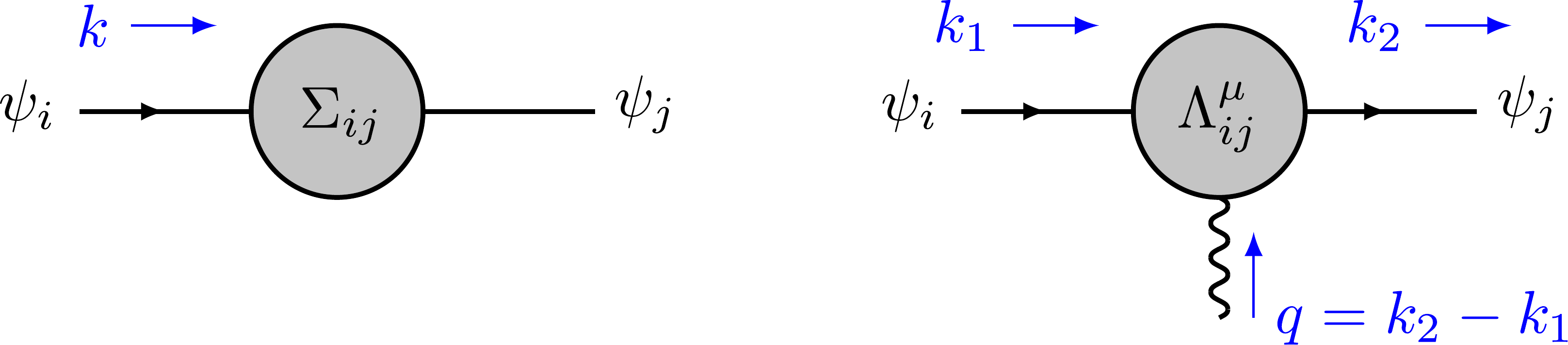}
\caption{Assignment of the momenta for the self-energy and vertex correction.
}
\label{fig:L_eff}
\end{center}
\end{figure}

The amplitude of each diagram in Fig.~\ref{fig:diagram_fermion_scalar} can be given by
\begin{align}
i\mathcal{M}_{\mathrm{A1}}=&~
eQ_{\psi}\int\!\frac{d^4k}{(2\pi)^4}
\overline{u}(p_2)P_R\left(\slashed{k}_2 + m_j\right)
\tilde{\Sigma}_{ji}(\slashed{k}_2)
\left(\slashed{k}_2 + m_i\right)
\slashed{\epsilon}\left(\slashed{k}_1+m_i\right)
P_Lu(p_1)\nonumber\\
&\times \frac{1}{(k-p)^2-m_{\phi}^2}\frac{1}{k_1^2-m_i^2}\frac{1}{k_2^2-m_i^2}\frac{1}{k_2^2-m_j^2},\\
i\mathcal{M}_{\mathrm{A2}}=&~
eQ_{\psi}\int\!\frac{d^4k}{(2\pi)^4}
\overline{u}(p_2)P_R\left(\slashed{k}_2+m_j\right)
\slashed{\epsilon}\left(\slashed{k}_1+m_j\right)
\tilde{\Sigma}_{ji}(\slashed{k}_1)
\slashed{\epsilon}\left(\slashed{k}_1+m_i\right)
P_Lu(p_1)\nonumber\\
&\times \frac{1}{(k-p)^2-m_{\phi}^2}\frac{1}{k_1^2-m_i^2}\frac{1}{k_1^2-m_j^2}\frac{1}{k_2^2-m_j^2},\\
i\mathcal{M}_{\mathrm{A3}}=&~
e\epsilon_{\mu}\int\!\frac{d^4k}{(2\pi)^4}
\overline{u}(p_2)P_R\left(\slashed{k}_2+m_j\right)
\tilde{\Lambda}_{ji}^{\mu}(k_1,k_2)
\left(\slashed{k}_1+m_i\right)
P_Lu(p_1)\nonumber\\
&\times \frac{1}{(k-p)^2-m_{\phi}^2}\frac{1}{k_1^2-m_i^2}\frac{1}{k_2^2-m_j^2},
\label{eq:amp3}\\
i\mathcal{M}_{\mathrm{A4}}=&~
eQ_{\phi}\epsilon_{\mu}\int\!\frac{d^4k}{(2\pi)^4}\overline{u}(p_2)P_R
\left(\slashed{k}+m_j\right)\tilde{\Sigma}_{ji}(\slashed{k})
\left(\slashed{k}+m_i\right)P_Lu(p_1)\nonumber\\
&\times\frac{1}{k^2-m_i^2}\frac{1}{k^2-m_j^2}
\frac{1}{(k-p_1)^2-m_{\phi}^2}\frac{1}{(k-p_2)^2-m_{\phi}^2}\left(p_1+p_2-2k\right)^{\mu},
\label{eq:amp4}
\end{align}
where $\tilde{\Sigma}_{ji}(\slashed{k})$ and $\tilde{\Lambda}_{ji}^{\mu}(k_1,k_2)$ are defined by
\begin{align}
\tilde{\Sigma}_{ji}(\slashed{k})\equiv y_{j\alpha}^*\Sigma_{ji}(\slashed{k}) y_{i\alpha},\qquad
\tilde{\Lambda}_{ji}^{\mu}(k_1,k_2)\equiv y_{j\alpha}^*\Lambda_{ji}^{\mu}(k_1,k_2)y_{i\alpha},
\label{eq:tilde}
\end{align}
and the momenta in the above equations are defined by $q\equiv k_2-k_1(=p_2-p_1)$, $k_1\equiv k-q/2$, $k_2\equiv k+q/2$ and $p\equiv (p_1+p_2)/2$ as shown in Fig.~\ref{fig:L_eff}.
The photon polarization vector is given by $\epsilon_{\mu}$. 
The fermion self-energy $\Sigma_{ji}(\slashed{k})$ can be expressed by 
\begin{align}
 \Sigma_{ji}(\slashed{k})=
A_{ji}^{L}(k^2)\slashed{k}P_L+A_{ji}^{R}(k^2)\slashed{k}P_R+B_{ji}^{L}(k^2)P_L+B_{ji}^{R}(k^2)P_R,
\label{eq:self-energy}
\end{align}
with the form factors $A_{ji}^{L/R}(k^2)$ and $B_{ji}^{L/R}(k^2)$ for each chirality. 
All of these form factors are Lorentz scalars as a function of $k^2$. 
This expansion is always possible without loss of generality at any order of perturbative expansions. 
We impose the hermiticity on the self-energy since the absorptive part of the self-energy has nothing to do with our EDM calculation.\footnote{This fact can be found by considering whether or not all the intermediate states can satisfy the on-shell conditions in the EDM loop diagrams. 
We focus on the situation where the total mass of loop propagating particles are heavier than the mass of the external fermion mass. As long as this condition is satisfied, all the intermediate states are not allowed to be on-shell. }
Namely, the condition
\begin{align}
  \big(\Sigma^\dagger\big)_{j  i}  \gamma^0  
  =  \gamma^0  \Sigma_{i  j},
\end{align}
is imposed, and thus we obtain the relations between the form factors. 
\begin{align}
A_{ji}^L=A_{ij}^{L*}, \qquad
A_{ji}^R=A_{ij}^{R*}, \qquad
B_{ji}^L=B_{ij}^{R*}, \qquad
B_{ji}^R=B_{ij}^{L*}. 
\label{eq:herm}
\end{align}
In the following calculation, we express the amplitude in terms of $A_{ji}^L$, $A_{ji}^R$ and $B_{ji}^L(\equiv B_{ji})$. 

\subsubsection{The amplitude (A1)+(A2)}
As already stated, it is sufficient to expand the amplitude up to $\mathcal{O}(q)$ in order to extract the term of static EDMs~\cite{Shabalin:1978rs}. 
The CP violating part of the partial amplitude $i\mathcal{M}_\mathrm{A1+A2}\equiv i\mathcal{M}_\mathrm{A1}+i\mathcal{M}_\mathrm{A2}$ 
relevant to the EDM of the fermion $f_\alpha$ can be evaluated as
\begin{align}
 i\mathcal{M}_\mathrm{A1+A2}^\mathrm{CP}\approx&~
2ieQ_{\psi}\int\!\frac{d^4k}{(2\pi)^4}
\overline{u}(p_2)\slashed{\epsilon}P_Lu(p_1)
\frac{\left(k\!\cdot\!q\right)m_i}{(k-p)^2-m_{\phi}^2}\frac{1}{\left(k^2-m_i^2\right)^2}\frac{1}{\left(k^2-m_j^2\right)^2}
\nonumber\\
&\times
\mathrm{Im}\Bigg[
k^4m_i
\frac{d\tilde{A}_{ji}^{R}}{dk^2}
-m_j^3\bigg(\tilde{A}_{ji}^{L}+k^2\frac{d\tilde{A}_{ji}^L}{dk^2}\bigg)
+\left(k^4-2k^2m_j^2+m_i^2m_j^2\right)\frac{d\tilde{B}_{ji}}{dk^2}\Bigg]\nonumber\\
&-2ieQ_{\psi}\int\!\frac{d^4k}{(2\pi)^4}
\overline{u}(p_2) \slashed{k} \slashed{\epsilon} \slashed{k} P_Lu(p_1)
\frac{\left(k\!\cdot\!q\right)m_j^2}{(k-p)^2-m_{\phi}^2}\frac{1}{\left(k^2-m_i^2\right)^2}\frac{1}{\left(k^2-m_j^2\right)^2}\nonumber\\
&\times\mathrm{Im}
\Bigg[\tilde{A}_{ji}^{R}+k^2\frac{d\tilde{A}_{ji}^{R}}{dk^2}+m_im_j\frac{d\tilde{A}_{ji}^L}{dk^2}
+2m_i\frac{d\tilde{B}_{ji}}{dk^2}\Bigg],
\label{eq:amp1+2_2}
\end{align}
where the hermiticity condition in Eq.~(\ref{eq:herm}) is used to reduce the equation. 
We introduce the following notation in the same manner as defined in Eq.~\eqref{eq:tilde}. 
\begin{align}
  \tilde{A}_{ji}^{L/R}(k^2)
  \equiv  y_{j\alpha}^*  A_{ji}^{L/R} (k^2)  y_{i\alpha},
  \qquad
  \tilde{B}_{ji} (k^2)
  \equiv  y_{j\alpha}^*  B_{ji} (k^2)  y_{i\alpha}.
\end{align}

\subsubsection{The amplitude (A3)}
For the amplitude $i\mathcal{M}_\mathrm{A3}$, 
the vertex correction $\Lambda^{\mu}(k_1,k_2)$ obeys the Ward-Takahashi identity
\begin{align}
q_{\mu}\Lambda^{\mu}(k_1,k_2)=\Sigma\left(\slashed{k}_1\right)-\Sigma\left(\slashed{k}_2\right).
\label{eq:wti}
\end{align}
The full vertex function $\Gamma^{\mu}$ involving the tree-level contribution is related with the correction $\Lambda^{\mu}$ as $\Gamma^{\mu}=\gamma^{\mu}+\Lambda^{\mu}$. 
The vertex correction $\Lambda^{\mu}(k_1,k_2)$ can be decomposed into the longitudinal part $\Lambda_\mathrm{L}^{\mu}(k_1,k_2)$ and 
the transverse part $\Lambda_\mathrm{T}^{\mu}(k_1,k_2)$, which is orthogonal to the photon momentum~\cite{Ball:1980ay}, namely
\begin{align}
\Lambda^{\mu}(k_1,k_2)=\Lambda_\mathrm{L}^{\mu}(k_1,k_2)+\Lambda_\mathrm{T}^{\mu}(k_1,k_2),
\end{align}
where $\Lambda_\mathrm{L}^{\mu}(k_1,k_2)$ and $\Lambda_\mathrm{T}^{\mu}(k_1,k_2)$ obey
\begin{align}
q_{\mu}\Lambda_\mathrm{L}^{\mu}(k_1,k_2)=\Sigma\left(\slashed{k}_1\right)-\Sigma\left(\slashed{k}_2\right),\qquad
q_{\mu}\Lambda_\mathrm{T}^{\mu}(k_1,k_2)=0.
\label{eq:wt}
\end{align}

In general, the vertex correction $\Lambda^{\mu}(k_1,k_2)$ can be expanded by 12 independent vectors 
(24 vectors taking into account chirality operators $P_L$ and $P_R$ for any chiral models)~\cite{Bernstein:1968aa, Ball:1980ay}. 
However, 4 of these vectors (8 for chiral models) can be eliminated by the Ward-Takahashi identity shown in Eq.~(\ref{eq:wti}).\footnote{The detailed calculation is given in Appendix~\ref{sec:append1}.} 
As a result, the longitudinal part is written as 
\begin{align}
\Lambda_\mathrm{L}^{\mu}(k_1,k_2)=&~
\Bigg[\frac{A^L(k_1^2)-A^L(k_2^2)}{\left(k\!\cdot\!q\right)}k^{\mu}\slashed{k}
-\frac{A^L(k_1^2)+A^L(k_2^2)}{2}\gamma^{\mu}
+\frac{B(k_1^2)-B(k_2^2)}{\left(k\!\cdot\!q\right)}k^{\mu}\Bigg]P_L\nonumber\\
&~+
\Bigg[\frac{A^R(k_1^2)-A^R(k_2^2)}{\left(k\!\cdot\!q\right)}k^{\mu}\slashed{k}
-\frac{A^R(k_1^2)+A^R(k_2^2)}{2}\gamma^{\mu}
+\frac{B^{\dag}(k_1^2)-B^{\dag}(k_2^2)}{\left(k\!\cdot\!q\right)}k^{\mu}\Bigg]P_R,
\label{eq:long}
\end{align}
with the form factors in Eq.~(\ref{eq:self-energy}).
On the other hand, the transverse part is given by the remaining 8 independent vectors $V_{a}^{\mu}~(a=1-8)$ as
\begin{align}
 \Lambda_\mathrm{T}^{\mu}(k_1,k_2)=&~
\sum_{a=1}^{8}\Big[C_{a}^{L}(k_1,k_2)V_{a}^{\mu}P_L+C_{a}^{R}(k_1,k_2)V_{a}^{\mu}P_R\Big],
\label{eq:trans}
\end{align}
where we take the independent vectors $V_a^{\mu}$ as follows
\begin{align}
V_1^{\mu}=&~\left(k\!\cdot\!q\right)q^{\mu}-q^2k^{\mu},\hspace{1.35cm}
V_2^{\mu}=\slashed{k}\left[\left(k\!\cdot\!q\right)q^{\mu}-q^2k^{\mu}\right],\nonumber\\
V_3^{\mu}=&~\slashed{k}\slashed{q}\left[\left(k\!\cdot\!q\right)q^{\mu}-q^2k^{\mu}\right],\hspace{0.5cm}
V_4^{\mu}=\slashed{q}q^{\mu}-q^2\gamma^{\mu},\nonumber\\
V_5^{\mu}=&~\left(k\!\cdot\!q\right)\gamma^{\mu}-\slashed{q}k^{\mu},\hspace{1.5cm}
V_6^{\mu}=\slashed{k}\left[\left(k\!\cdot\!q\right)\gamma^{\mu}-\slashed{q}k^{\mu}\right],\nonumber\\
V_7^{\mu}=&~i\sigma^{\mu\nu}q_{\nu},\hspace{3.0cm}
V_8^{\mu}=i\epsilon^{\mu\nu\rho\sigma}\gamma_{5}\gamma_{\nu}k_{\rho}q_{\sigma},
\end{align}
with $\sigma^{\mu\nu}\equiv i\left(\gamma^{\mu}\gamma^{\nu}-\gamma^{\nu}\gamma^{\mu}\right)/2$ and the anti-symmetric tensor $\epsilon^{\mu\nu\rho\sigma}~(\epsilon^{0123}=+1)$. 
Note that the form factors $C_a^L$ and $C_a^R$ can be regarded as a function of only $k^2$, namely $C_a^L(k^2)$ and $C_a^R(k^2)$, 
as long as the $\mathcal{O}(q)$ amplitude is considered because the vectors $V_a^{\mu}$ are already $\mathcal{O}(q)$ at least.
One can explicitly show that Eqs.~(\ref{eq:long}) and (\ref{eq:trans}) obey Eq.~(\ref{eq:wt}). 
Furthermore, the following relations are satisfied,
\begin{align}
\frac{\partial\Lambda_\mathrm{L}^\mu}{\partial q_{\nu}}\Big|_{q=0}=0,\qquad
\Lambda_\mathrm{T}^{\mu}(k,k)=0.
\end{align}

From the above argument, the vertex correction expanded up to $\mathcal{O}(q)$ can be separated into the derivative of the self-energy and the transverse part as follows~\cite{Cvitanovic:1974um}, 
\begin{align}
\Lambda^{\mu}(k_1,k_2)\approx&~
\Lambda^{\mu}\Big|_{q=0}
+q_{\nu}\frac{\partial\Lambda^{\mu}}{\partial q_{\nu}}\Big|_{q=0}\nonumber\\
=&~
\Lambda_\mathrm{L}^{\mu}\Big|_{q=0}
+q_{\nu}\frac{\partial\Lambda_\mathrm{T}^{\mu}}{\partial q_{\nu}}\Big|_{q=0}
=
-\frac{d\Sigma}{dk_{\mu}}+q_{\nu}\frac{\partial\Lambda_\mathrm{T}^{\mu}}{\partial q_{\nu}}\Big|_{q=0}.
\label{eq:lambda}
\end{align}
Using Eq.~(\ref{eq:lambda}), the amplitude $i\mathcal{M}_\mathrm{A3}$ can be evaluated at $\mathcal{O}(q)$. 
The CP violating part coming from the first term of the last expression in Eq.~(\ref{eq:lambda}), 
which corresponds to the longitudinal part of the vertex correction, 
is calculated as
\begin{align}
 i\mathcal{M}_{\mathrm{A3L}}^\mathrm{CP}\approx&
-2ieQ_{\psi}\int\!\frac{d^4k}{(2\pi)^4}\overline{u}(p_2)\slashed{\epsilon}P_Lu(p_1)\frac{\left(k\!\cdot\!q\right)m_i}{(k-p)^2-m_{\phi}^2}
\frac{1}{\left(k^2-m_i^2\right)^2}\frac{1}{\left(k^2-m_j^2\right)^2}\nonumber\\
&\times\mathrm{Im}\Bigg[
k^4m_i\frac{d\tilde{A}_{ji}^{R}}{dk^2}-m_j^3\Bigg(\tilde{A}_{ji}^{L}+k^2\frac{d\tilde{A}_{ji}^{L}}{dk^2}\Bigg)
+k^2\left(m_i^2-m_j^2\right)\frac{d\tilde{B}_{ji}}{dk^2}
\Bigg]\nonumber\\
&+2ieQ_{\psi}\int\!\frac{d^4k}{(2\pi)^4}
\overline{u}(p_2) \slashed{k} \slashed{\epsilon} \slashed{k} P_Lu(p_1)
\frac{\left(k\!\cdot\!q\right)m_j^2}{(k-p)^2-m_{\phi}^2}\frac{1}{\left(k^2-m_i^2\right)^2}\frac{1}{\left(k^2-m_j^2\right)^2}\nonumber\\
&\times\mathrm{Im}
\Bigg[\tilde{A}_{ji}^{R}+k^2\frac{d\tilde{A}_{ji}^{R}}{dk^2}+m_im_j\frac{d\tilde{A}_{ji}^L}{dk^2}
+2m_i\frac{d\tilde{B}_{ji}}{dk^2}\Bigg]\nonumber\\
&-2ieQ_{\psi}\int\!\frac{d^4k}{(2\pi)^4}\overline{u}(p_2)\slashed{q}P_Lu(p_1)\frac{\left(\epsilon\!\cdot\!k\right)m_i}{(k-p)^2-m_{\phi}^2}\frac{1}{k^2-m_i^2}\frac{1}{k^2-m_j^2}
\mathrm{Im}\Bigg[\frac{d\tilde{B}_{ji}}{dk^2}\Bigg],
\label{eq:amp3_2}
\end{align}
where the hermiticity condition given in Eq.~(\ref{eq:herm}) and the conservation of the electromagnetic charges $\sum_{p}Q_{p}=Q_{\psi}$ are used. 

The remaining transverse part of the amplitude $i\mathcal{M}_\mathrm{A3}$, which corresponds to the second term of the last expression in Eq.~(\ref{eq:lambda}), can be evaluated. 
First, since the vectors $V_{a}^{\mu}~(a=1-4)$ are $\mathcal{O}(q^2)$, we can ignore these terms in the EDM calculation. 
Second, one can find that the terms for the vectors $V_5^{\mu}$ and $V_8^{\mu}$ have no CP violation as discussed below. 
The following relations are derived from the hermiticity of the vertex correction $\Lambda_\mathrm{T}^{\mu}(k_1,k_2)$
\begin{align*}
 \left(C_{5}^L\right)_{ji}=\left(C_{5}^L\right)_{ij}^*,\qquad
\left(C_{5}^R\right)_{ji}=\left(C_{5}^R\right)_{ij}^*,\\
 \left(C_{6}^L\right)_{ji}=\left(C_{6}^R\right)_{ij}^*,\qquad
\left(C_{6}^R\right)_{ji}=\left(C_{6}^L\right)_{ij}^*,\\
 \left(C_{7}^L\right)_{ji}=\left(C_{7}^R\right)_{ij}^*,\qquad
\left(C_{7}^R\right)_{ji}=\left(C_{7}^L\right)_{ij}^*,\\
 \left(C_{8}^L\right)_{ji}=\left(C_{8}^L\right)_{ij}^*,\qquad
\left(C_{8}^R\right)_{ji}=\left(C_{8}^R\right)_{ij}^*.
\end{align*}
Although each term is complex, multiplying the Yukawa coupling $y_{j\alpha}^*$ and $y_{i\alpha}$, and summing up $i$ and $j$, 
it can be found that only the real part remains and the imaginary part involving CP violation vanishes for the vectors $V_5^{\mu}$ and $V_8^{\mu}$.
On the other hand, the terms for the vectors $V_6^{\mu}$ and $V_7^{\mu}$ involving even numbers of gamma matrices violate the CP. 
From the above consideration, one can extract the CP violating term from the transverse part as 
\begin{align}
i\mathcal{M}_{\mathrm{A3T}}^\mathrm{CP}=&~
2iem_i\int\!\frac{d^4k}{(2\pi)^4}
\overline{u}(p_2)\Big[\left(k\!\cdot\!q\right)\slashed{\epsilon}-\left(\epsilon\!\cdot\!k\right)\slashed{q}\Big]P_Lu(p_1)\nonumber\\
&\times\frac{1}{(k-p)^2-m_{\phi}^2}\frac{1}{k^2-m_i^2}\frac{1}{k^2-m_j^2}
\mathrm{Im}\bigg[
k^2\tilde{C}_{ji}(k^2)+\tilde{D}_{ji}(k^2)
\bigg],
\label{eq:3T}
\end{align}
where $\tilde{C}_{ji}(k^2)$ and $\tilde{D}_{ji}(k^2)$ are defined by the sum of all the vertex corrections for the left-chirality
\begin{align}
\tilde{C}_{ji}(k^2)\equiv \sum_{p}Q_p\Big[y_{j\alpha}^*\big(C_{p6}^{L}(k^2)\big)_{ji}y_{i\alpha}\Big],\quad
\tilde{D}_{ji}(k^2)\equiv \sum_{p}Q_p\Big[y_{j\alpha}^*\big(C_{p7}^{L}(k^2)\big)_{ji}y_{i\alpha}\Big].
\label{eq:cd}
\end{align}

\subsubsection{The amplitude (A4)}
The amplitude $i\mathcal{M}_\mathrm{A4}$ can be evaluated by directly substituting the self-energy of Eq.~(\ref{eq:self-energy}) into Eq.~(\ref{eq:amp4})
\begin{align}
i\mathcal{M}_\mathrm{A4}=&~
eQ_{\phi}\int\!\frac{d^4k}{(2\pi)^4}
\overline{u}(p_2)\slashed{k}P_Lu(p_1)
\Big[\epsilon\!\cdot\!\left(p_1+p_2-2k\right)\Big]\nonumber\\
&~\times
\Big[m_jm_i\tilde{A}_{ji}^L(k^2)+k^2\tilde{A}_{ji}^R(k^2)+m_i\tilde{B}_{ji}(k^2)+m_j\tilde{B}_{ij}^*(k^2)\Big]\nonumber\\
&~\times\frac{1}{k^2-m_i^2}\frac{1}{k^2-m_j^2}\frac{1}{(k-p_1)^2-m_{\phi}^2}\frac{1}{(k-p_2)^2-m_{\phi}^2}.
\label{eq:amp4_2}
\end{align}
Note that the indices $i$ and $j$ are summed up. 
One can find that the CP violating part is eventually zero due to the hermiticity of the self-energy, and thus there is no contribution to the EDM in the amplitude.

\subsubsection{The total amplitude}
Combining Eqs.~(\ref{eq:amp1+2_2}), (\ref{eq:amp3_2}) and (\ref{eq:3T}), the CP violating term of the total amplitude in Fig.~\ref{fig:diagram_fermion_scalar} is summarized as
\begin{align}
 i\mathcal{M}_\mathrm{scalar}^\mathrm{CP}=&~2iem_i\int\!\frac{d^4k}{(2\pi)^4}
\overline{u}(p_2)\left[
\left(k\!\cdot\!q\right)\slashed{\epsilon}-\left(\epsilon\!\cdot\!k\right)\slashed{q}
\right]P_Lu(p_1)
\frac{1}{\left(k-p\right)^2-m_{\phi}^2}\frac{1}{k^2-m_i^2}\frac{1}{k^2-m_j^2}\nonumber\\
&
\times\mathrm{Im}\left[
Q_{\psi}\frac{d\tilde{B}_{ji}}{d k^2}
+k^2\tilde{C}_{ji}(k^2)+\tilde{D}_{ji}(k^2)
\right].
\label{eq:result}
\end{align}
This is our main result of the reduction of the EDM calculation. 
Note that we do not use any approximation to derive Eq.~(\ref{eq:result}) except for the $\mathcal{O}(q)$ expansion, which does not affect the EDM calculation. 
Therefore, the above calculation for the EDM is exact. 

The formula in Eq.~(\ref{eq:result}) does not directly give the EDM of the fermion $f_\alpha$, and 
we have to assume a specific model to further proceed with the calculation to obtain a concrete formula of EDMs. 
However, once a model is fixed, using the Gordon identity, the EDM coefficient $d_\alpha$ can straightforwardly be extracted from
\begin{align}
 i\mathcal{M}_\mathrm{total}^\mathrm{CP}=
d_{\alpha}\epsilon_{\mu}\overline{u}(p_2)i\sigma^{\mu\nu}q_{\nu}\gamma_5u(p_1),
\label{eq:extract}
\end{align}
where $q=p_2-p_1$. 
In particular, if $p^2=m_{\alpha}^2\ll m_{\phi}^2$, one can expand the $\phi$ propagator in Eq.~(\ref{eq:result}) as
\begin{align}
\frac{1}{(k-p)^2-m_\phi^2}&\approx\frac{1}{k^2-m_\phi^2}+\frac{2\left(k\!\cdot\!p\right)}{\left(k^2-m_\phi^2\right)^2}.
\label{eq:expand}
\end{align}
Then it can be found that the contribution from the first term in the right-hand side of 
Eqs.~(\ref{eq:expand}) vanishes because the resultant integrand is an odd function 
in terms of the loop momentum $k$. 
Thus the second term in Eq.~(\ref{eq:expand}) gives a leading contribution to the EDM. 
One can derive the approximate EDM formula for $m_\alpha^2\ll m_\phi^2$,
\begin{align}
\frac{d_\alpha}{e}\approx&
-\frac{im_im_\alpha}{2}\int\!\frac{d^4k}{(2\pi)^4}
\frac{k^2}{\left(k^2-m_{\phi}^2\right)^2}\frac{1}{k^2-m_i^2}\frac{1}{k^2-m_j^2}\nonumber\\
&\times
\mathrm{Im}\left[
Q_{\psi}\frac{d\tilde{B}_{ji}}{d k^2}
+k^2\tilde{C}_{ji}(k^2)+\tilde{D}_{ji}(k^2)
\right],
\label{eq:result_app}
\end{align}
where $\tilde{B}_{ji}(k^2)$ is the form factor of the self-energy, and $\tilde{C}_{ji}(k^2)$ and $\tilde{D}_{ji}(k^2)$ are defined in Eq.~(\ref{eq:cd}).\footnote{We can read out the higher dimensional operator that induces the EDM with equations of motion. 
Here, we introduce the following loop
function.
\begin{align}
  &
  \int\!\frac{d^4k}{(2\pi)^4}
  \frac{k_\mu}{\left[ \left(k-p\right)^2-m_{\phi}^2 \right]  \left( k^2-m_i^2 \right)  \left( k^2-m_j^2 \right)}
  \mathrm{Im}
  \left[
    Q_{\psi}\frac{d\tilde{B}_{ji}}{d k^2}+k^2\tilde{C}_{ji}(k^2)+\tilde{D}_{ji}(k^2)
  \right]
  \equiv
  i  C^{\rm  CP}  p_\mu,
  \nonumber
\end{align}
where $C^{\rm  CP}$ is the scalar function. 
After performing the loop integral of Eq.~\eqref{eq:result}, 
we can read out the dimension six operator shown below. 
\begin{align}
  {\cal  L}_{\rm  eff} 
  =  
  -  e  m_i  C^{\rm  CP}
  \left[  
    \overline{f}_\alpha  i  {D^\mu\raisebox{0.33cm}{\tiny{$\hspace{-0.45cm}\leftrightarrow$}\hspace{0.1cm}}}  ~\gamma^\nu  P_L  f_\alpha
  \right]  F_{\mu  \nu}, 
  \nonumber
\end{align}
where $D_\mu$ is the covariant derivative for $f_\alpha$.
}

In the next section, some examples of the EDM calculation using Eq.~(\ref{eq:result}) will be shown in specific models.

\begin{figure}[t]
\begin{center}
\includegraphics[scale=1.1]{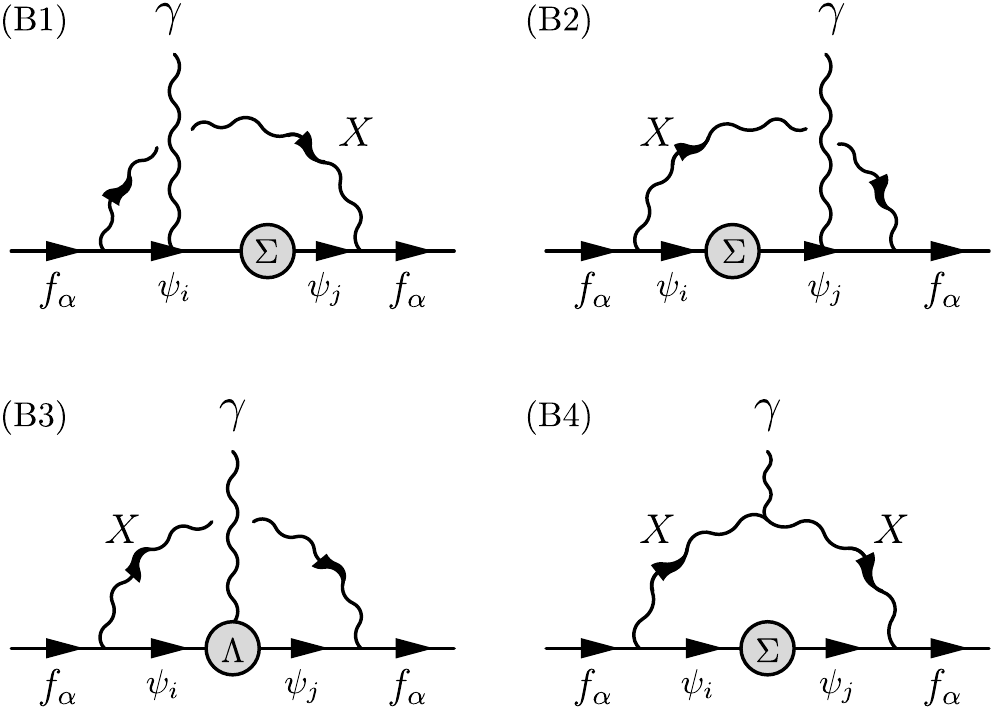}
\caption{Feynman diagrams potentially inducing non-zero EDMs of the fermion $f_\alpha$ with the multi-fermions $\psi_i$ and a vector boson $X$.}
\label{fig:diagram_fermion_vector}
\end{center}
\end{figure}

\subsubsection{EDM reduction for diagrams with vector boson in outer loop}
\label{sec:vector_loop}
For the case that a vector boson propagates instead of the scalar $\phi$ as shown in Fig.~\ref{fig:diagram_fermion_vector}, the calculation is essentially the same as the scalar case. 
The Lagrangian for this case is given by
\begin{align}
\mathcal{L}=&-\Big(g_{i\alpha}X_{\mu}^{*}\overline{\psi_i}\gamma^{\mu}P_Lf_\alpha+\mathrm{H.c.}\Big)
-\overline{\psi}_i\Sigma_{ij}\psi_j-eA_{\mu}\overline{\psi_i}\Lambda_{ij}^{\mu}\psi_j\nonumber\\
&-Q_{\psi}A_{\mu}\overline{\psi_i}\gamma^{\mu}\psi_i
+ieQ_X\Big(
A_{\mu}\big(X_{\nu}^{*}\partial^{\mu}\raisebox{0.33cm}{\tiny{$\hspace{-0.4cm}\leftrightarrow$}\hspace{0.1cm}}X^{\nu}\big)
+X_{\mu}\big(A_{\nu}\partial^{\mu}\raisebox{0.33cm}{\tiny{$\hspace{-0.4cm}\leftrightarrow$}\hspace{0.1cm}}X^{*\nu}\big)
+X_{\mu}^{*}\big(X_{\nu}\partial^{\mu}\raisebox{0.33cm}{\tiny{$\hspace{-0.4cm}\leftrightarrow$}\hspace{0.1cm}}A^{\nu}\big)
\Big).
\end{align}
There is no contribution to EDM from the diagram (B4), and the CP violating term induced by the other diagrams is summarized as
\begin{align}
 i\mathcal{M}_\mathrm{vector}^\mathrm{CP}=&~-2iem_i\int\!\frac{d^4k}{(2\pi)^4}
\overline{u}(p_2)\gamma^{\rho}\Big[
\left(k\!\cdot\!q\right)\slashed{\epsilon}-\left(\epsilon\!\cdot\!k\right)\slashed{q}
\Big]P_L\gamma^{\sigma}u(p_1)\nonumber\\
&~\times\frac{1}{\left(k-p\right)^2-m_{X}^2}\frac{1}{k^2-m_i^2}\frac{1}{k^2-m_j^2}
\left(g_{\rho\sigma}-\frac{\left(k-p\right)_{\rho}\left(k-p\right)_{\sigma}}{m_X^2}\right)\nonumber\\
&~\times\mathrm{Im}\left[
Q_{\psi}\frac{d \tilde{B}_{ji}}{d k^2}
+k^2\tilde{C}_{ji}(k^2)+\tilde{D}_{ji}(k^2)
\right],
\label{eq:result2}
\end{align}
where $p\equiv(p_1+p_2)/2$ as same as the scalar case, and we chose the unitary gauge.
The form factors $\tilde{B}(k^2)$, $\tilde{C}(k^2)$ and $\tilde{D}(k^2)$ are defined as same as Eqs.~\eqref{eq:tilde}~and~\eqref{eq:cd} with the gauge coupling $g_{j\alpha}^*$, $g_{i\alpha}$ 
instead of the Yukawa coupling $y_{j\alpha}^*$, $y_{i\alpha}$. 

With the approximation $p^2=m_\alpha^2\ll m_X^2$, the EDM formula can be extracted from Eq.~(\ref{eq:extract}), which is given by\footnote{To read out the higher dimensional operators, we introduce the following loop functions.
\begin{align}
  &
  \int\!\frac{d^4k}{(2\pi)^4}
  \frac{k_\mu  \left(g_{\rho\sigma}-\frac{\left(k-p\right)_{\rho}\left(k-p\right)_{\sigma}}{m_X^2}\right)}{\left[ \left(k-p\right)^2-m_{X}^2 \right]  \left( k^2-m_i^2 \right)  \left( k^2-m_j^2 \right)}
  \mathrm{Im}
  \left[
    Q_{\psi}\frac{d\tilde{B}_{ji}}{d k^2}+k^2\tilde{C}_{ji}(k^2)+\tilde{D}_{ji}(k^2)
  \right]
  \nonumber
  \\
  \equiv&~
  i  C^{\rm  CP}_1  p_\mu  g_{\rho  \sigma}
  +  i  C^{\rm  CP}_2  \left( g_{\mu  \rho}  p_\sigma +  g_{\mu  \sigma}  p_\rho \right)
  +  i  C^{\rm  CP}_3  p_\mu  p_\rho  p_\sigma,
  \nonumber
\end{align}
where $C^{\rm  CP}_i  (i  =  1,  2,  3)$ are the scalar functions. 
After performing the loop integral of Eq.~\eqref{eq:result2}, 
we can read out the dimension six and dimension eight operators shown below. 
\begin{align}
  {\cal  L}_{\rm  eff}
  =&  
  -  2  e  m_i  \left( C^{\rm  CP}_1  +  2  C^{\rm  CP}_2 \right)
  \left[  
    \overline{f}_\alpha  i  {D^\mu\raisebox{0.33cm}{\tiny{$\hspace{-0.45cm}\leftrightarrow$}\hspace{0.1cm}}}  ~\gamma^\nu  P_R  f_\alpha
  \right]  
  F_{\mu  \nu}
  +  \frac{e  m_i}{4}  C^{\rm  CP}_3
  \left[  
    \overline{f}_\alpha  i  {D^\mu\raisebox{0.33cm}{\tiny{$\hspace{-0.45cm}\leftrightarrow$}\hspace{0.1cm}}}  ~\gamma^\nu  \bigl( i {D\raisebox{0.33cm}{\tiny{$\hspace{-0.25cm}\leftrightarrow$}\hspace{0.1cm}}} \bigr)^2  P_R  f_\alpha
  \right]  
  F_{\mu  \nu}.
  \nonumber
\end{align}
These operators induce the EDMs with equations of motion.
}
\begin{align}
\frac{d_\alpha}{e}\approx&~\frac{3im_im_\alpha}{4}\int\!\frac{d^4k}{(2\pi)^4}
\frac{k^2}{m_X^2}\frac{k^2}{\left(k^2-m_X^2\right)^2}\frac{1}{k^2-m_i^2}\frac{1}{k^2-m_j^2}\nonumber\\
&\times\mathrm{Im}\left[
Q_{\psi}\frac{d \tilde{B}_{ji}}{d k^2}
+k^2\tilde{C}_{ji}(k^2)+\tilde{D}_{ji}(k^2)
\right].
\label{eq:result_app2}
\end{align}

As shown in our reduction formulas, Eq.~\eqref{eq:result} and Eq.~\eqref{eq:result2}, 
it is essential to pick up the chirality flipping effects from the internal loop 
in order to obtain the non-zero EDM contributions. 
We immediately conclude that EDM cancellation happens as long as we consider the setup where no chirality flipping effects are induced in the  self-energy and the vertex correction of the internal fermion.

\subsection{Reduction formulas for Chromo EDM}

We can apply our reduction scheme to the calculation of the chromo-EDM.
The chromo-EDM arises from the diagrams with the external gluon instead of the external photon. 
For the non-Abelian gauge theory, 
we have the Slavnov-Taylor identity imposed by the Becchi-Rouet-Stora-Tyutin (BRST) symmetry, 
which has a more complicated form than the Ward-Takahashi identity in the Abelian gauge theory. 
However, the Ward-Takahashi identity holds for the background gluon vertices, which is shown using the background gauging method~\cite{Abbott:1980hw,1405.3506}.
We can treat gluons as the background fields in the chromo-EDM calculation. 
We also note that the diagrams that the gluon couples to the outer loop do not induce the chromo-EDM for the same reason as the diagram (A4) and diagram (B4) in Fig.~\ref{fig:diagram_fermion_scalar} and Fig.~\ref{fig:diagram_fermion_vector}, respectively. 
The remaining diagrams have the same group factors of SU(3)$_c$ symmetry, 
which are automatically determined once we fix the concrete model, 
and thus the same reduction mechanism works. 
Consequently, 
we obtain the same EDM reduction formulas for the chromo-EDM except for the overall group factor.

\section{Application to specific models}
\label{sec:3}

In this section, 
we apply our reduction formulas to several models including the SM. 
As we clarify the conditions to get non-zero EDMs, 
we understand when and why we have exact EDM cancellation in rainbow-type diagrams.
We also show the setup that we can apply our formulas and obtain the non-zero values of the EDM prediction.

\subsection{The Standard Model}
The well-known result of the SM, 
in which the neutron EDM induced from the two-loop $W$ boson diagrams totally cancels out, 
can be reproduced from our formula in Eq.~(\ref{eq:result2}). 
Recall that only the left-handed fermions couple to the $W$ boson in the SM as in Eq.~(\ref{eq:lag_sm}). 
At the one-loop level, the self-energy for the up-type quarks is explicitly given by
\begin{align} 
  \Sigma_{j  i}  ( \slashed{k} )
  &=
  -  \frac{ig_2^2}{2} V_{j  \alpha} V_{i  \alpha}^* 
  \int\!
  \frac{d^4  \ell}{( 2  \pi )^4}
  \gamma^\rho
    P_L
    \frac{\slashed{k}-\slashed{\ell}+m_\beta}{( k-\ell )^2 - m_\beta^2}
    P_R
  \gamma^\sigma
    \frac{1}{\ell^2  -  m_W^2}
  \left( g_{\rho  \sigma}  -  \frac{\ell_\rho  \ell_\sigma}{m_W^2} \right).
\end{align}
Due to the projection operators $P_L$ and $P_R$, we have no chirality flipping term.
In this case, therefore, we only have $A^L_{j  i}  ( k^2 )$ 
in terms of the notation introduced in Eq.~\eqref{eq:self-energy}. 
Thus one can find that there is no EDM contribution from the self-energy. 

Regarding the contribution from the vertex correction $\Lambda_{ji}^{\mu}$, 
there are two possibilities that the photon can be attached to the down-type quark and the $W$ boson, which are explicitly given by
\begin{align}
\left(\Lambda_d\right)_{ji}^{\mu}(k_1,k_2)=&
-\frac{ig_2^2}{2}V_{j\beta}V_{i\beta}^{*}
\int\!\frac{d^4\ell}{(2\pi)^4}
\gamma^{\rho}P_L\frac{\left(\slashed{\ell}_2+m_{\beta}\right)}{\ell_2^2-m_\beta^2}
\gamma^{\mu}\frac{\left(\slashed{\ell}_1+m_\beta\right)}{\ell_1^2-m_\beta^2}
P_R\gamma^{\sigma}\nonumber\\
&~\times
\frac{1}{\left(k-\ell\right)^2-m_W^2}
\left(g_{\rho\sigma}-\frac{\left(k-\ell\right)_{\rho}\left(k-\ell\right)_{\sigma}}{m_W^2}\right),\\
\left(\Lambda_{W^+}\right)_{ji}^{\mu}(k_1,k_2)=&
-\frac{ig_2^2}{2}V_{j\beta}V_{i\beta}^*
\int\!\frac{d^4\ell}{(2\pi)^4}
\gamma^{\rho}P_L\frac{\left(\slashed{k}-\slashed{\ell}+m_\beta\right)}{\left(k-\ell\right)^2-m_\beta^2}P_R\gamma^{\sigma}
\frac{1}{\ell_1^2-m_W^2}\frac{1}{\ell_2^2-m_W^2}\nonumber\\
&~\times\Bigg[
+\left(g_{\sigma}^{~\nu}-\frac{\ell_{1\sigma}\ell_1^{\nu}}{m_W^2}\right)\left(g_{\nu\rho}-\frac{\ell_{2\nu}\ell_{2\rho}}{m_W^2}\right)
\left(\ell_1+\ell_2\right)^{\mu}\nonumber\\
&~\hspace{0.75cm}
-\left(g_{\sigma}^{~\mu}-\frac{\ell_{1\sigma}\ell_1^{\mu}}{m_W^2}\right)\left(g_{\nu\rho}-\frac{\ell_{2\nu}\ell_{2\rho}}{m_W^2}\right)
\left(\ell_1-q\right)^{\nu}\nonumber\\
&~\hspace{0.75cm}
-\left(g_{\sigma\nu}-\frac{\ell_{1\sigma}\ell_{1\nu}}{m_W^2}\right)
\left(g^{\mu}_{~\rho}-\frac{\ell_2^{\mu}\ell_{2\rho}}{m_W^2}\right)
\left(\ell_2+q\right)^{\nu}
\Bigg],
\end{align}
where $\ell_1=\ell-q/2$ and $\ell_2=\ell+q/2$. 
From these expressions, one can easily find that there is no contribution to the form factors $\tilde{C}_{ji}(k^2)$ and $\tilde{D}_{ji}(k^2)$ shown in Eq.~(\ref{eq:result2})
because only the odd numbers of the gamma matrices are involved in the vertex corrections.
Therefore the neutron EDM induced from the weak interactions in the SM at the two-loop level exactly vanishes, which is consistent with the result in Ref.~\cite{Shabalin:1978rs}.

\subsection{The scotogenic model}
In the scotogenic model proposed by E.~Ma~\cite{Ma:2006km}, it is known that the charged lepton EDMs induced from a subset of the diagrams exactly cancel~\cite{Abada:2018zra,Fujiwara:2020unw}. 
As in the SM, we have no chirality flipping term in the self-energy due to the chirality projection operators at the one-loop level. 
In addition, there is no $\tilde{C}_{ji}(k^2)$ and $\tilde{D}_{ji}(k^2)$ terms in Eq.~(\ref{eq:result}) coming from the vertex correction.
Therefore, the EDM cancellations, 
which have been found in the previous analysis~\cite{Abada:2018zra,Fujiwara:2020unw}, 
can be understood from our generalized EDM formula. 
Moreover, our derivation using form factors can be applicable for both Dirac and Majorana fermions.

On the other hand in this model, 
there are some additional two-loop diagrams that  are not categorized as the diagrams as shown in Fig.~\ref{fig:diagram_fermion_scalar}.\footnote{See diagrams shown in Fig.~2 of Ref.~\cite{Fujiwara:2020unw}, for example. We can not find the sub-diagram structures from these diagrams. }
Our calculation is not applicable for these diagrams, and indeed a non-zero contribution to charged lepton EDMs arises.

\subsection{The singlet-triplet extended model}

The singlet-triplet model which is named the ST model in Ref.~\cite{Fujiwara:2020unw} has been considered in the previous work. 
It has turned out that the charged lepton EDM arising from the rainbow-type diagrams also vanishes in this model.
Here we further extend the model introducing a triplet scalar $\Phi$ to demonstrate how we can obtain non-zero EDM from the rainbow-type diagrams using our formulas.
The particle contents are given in Tab.~\ref{tab:particles}, and the Lagrangian relevant to the charged lepton EDMs is given by
\begin{align}
\mathcal{L}=&~
\left(D_{\mu}\eta\right)^{\dag}\left(D^{\mu}\eta\right)+\frac{1}{2}\left(D_{\mu}\Phi\right)\left(D^{\mu}\Phi\right)
+\frac{1}{2}\overline{\Psi_i}\left(i\slashed{D}-m_i\right)\Psi_i\nonumber\\
&-\left(y_{i\alpha}\eta\overline{\Psi_i}P_LL_\alpha
+\frac{\lambda_{ij}}{2}\Phi\overline{\Psi_i^c}P_R\Psi_j+\mathrm{h.c.}\right),
\end{align}
where $\Psi_1\equiv N$ (singlet) and $\Psi_2\equiv\Sigma$ (triplet) and $i,j=1,2$. 
The new coupling $\lambda_{ij}$ which is not included in the original singlet-triplet model plays an important role in inducing a non-zero EDM due to chirality flipping effects.
Note that the cubic term $\Phi|H|_{\bm{3}}^2+\mathrm{h.c.}$ in the scalar potential induces the mixing between $H$ and $\Phi$ where $|H|_{\bm{3}}^2$ represents the $SU(2)_L$ triplet. 
However, this is ignored in the following calculation for simplicity.

\begin{table}[t]
\begin{center}
\begin{tabular}{|c||c|c|c|c|}\hline
 & $\eta$ & $\Phi$ & $N$ & $\Sigma$\\\hhline{|=#=|=|=|=|}
$SU(2)_L$ & $\bm{2}$ & $\bm{3}$ & $\bm{1}$ & $\bm{3}$\\\hline
$U(1)_Y$ & $1/2$ & $0$ & $0$ & $0$\\\hline
$\mathbb{Z}_2$ & $-1$ & $+1$ & $-1$ & $-1$\\\hline
Spin & $0$ & $0$ & $1/2$ & $1/2$\\\hline
\end{tabular}
\caption{Contents of the new particle in the singlet-triplet fermion model.}
\label{tab:particles}
\end{center}
\end{table}

\begin{figure}[t]
\begin{center}
\includegraphics[scale=0.77]{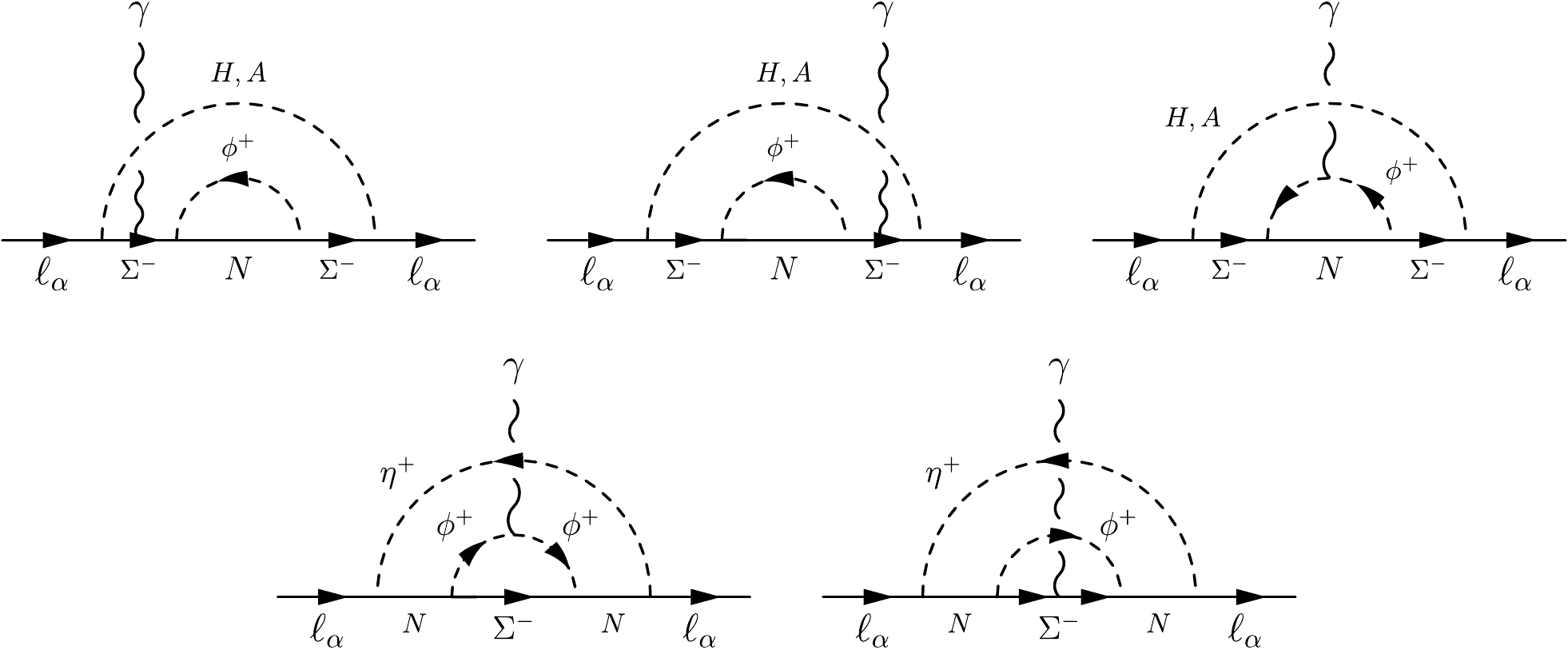}
\caption{Feynman diagrams inducing non-zero charged lepton EDM in the singlet-triplet extended model where $\eta^+$, $H$ and $A$ are the charged, CP-even and CP-odd neutral components of the $SU(2)_L$ doublet $\eta$, respectively.}
\label{fig:diagram_ex}
\end{center} 
\end{figure}

We can use our formula Eq.~(\ref{eq:result}) to extract the charged lepton EDM. 
The relevant Feynman diagrams are shown in Fig.~\ref{fig:diagram_ex}. 
For the three diagrams in the top, the EDM contribution comes from the self-energy correction (or equivalent to the longitudinal part of the vertex correction). 
The form factor $B_{ji}(k^2)$ of the self-energy can be evaluated as
\begin{align}
 B_{22}(k^2)=
-\frac{\lambda^{*2}m_1}{(4\pi)^2}\int_{0}^{1}\!dxdy\delta(x+y-1)
\lim_{\epsilon\to0}\Gamma\left(\frac{\epsilon}{2}\right)\left(\frac{4\pi}{\Delta_1(k^2)}\right)^{\epsilon/2},
\end{align}
where $\Delta_1(k^2)\equiv xm_1^2+ym_{\phi^+}^2-xyk^2$, $\lambda\equiv\left(\lambda_{12}+\lambda_{21}\right)/2$ and $\epsilon\equiv4-d$ is the parameter of the dimensional regularization.
There is no contribution from the transverse part of the vertex correction since it does not include terms with two gamma matrices.

For the two diagrams at the bottom of Fig.~\ref{fig:diagram_ex}, the EDM contribution is induced from the vertex correction of the right diagram, which is evaluated as 
\begin{align}
D_{11}(k^2)=
\frac{\lambda^{*2}}{(4\pi)^2}\int_{0}^{1}dxdy\delta(x+y-1)\frac{xm_2}{\Delta_2(k^2)},
\end{align}
where $\Delta_2(k^2)=xm_2^2+ym_{\phi^+}^2-xyk^2$.
Substituting the above expressions into the formula of Eq.~(\ref{eq:result}) and using the Gordon identity, we can extract the EDM from Eq.~(\ref{eq:extract}). 
The result is summarized as
\begin{align}
\frac{d_{\alpha}}{e}=&~
\frac{\mathrm{Im}\lambda^2}{2(4\pi)^4}
\frac{m_{\alpha}}{m_{\eta^+}^2}\sqrt{\xi_1\xi_2}
\int_{0}^{1}\!dxdy\delta(x+y-1)
\int_{0}^{1}\!dsdtdu\delta(s+t+u-1)\nonumber\\
&\times xy^2su\Bigg[
\frac{x|y_{2\alpha}|^2}{\Delta^2\left(\xi_2,\xi_1,\xi_H,\xi_{\phi^+},\xi_\alpha\right)}
+\frac{x|y_{2\alpha}|^2}{\Delta^2\left(\xi_2,\xi_1,\xi_A,\xi_{\phi^+},\xi_\alpha\right)}
-\frac{2|y_{1\alpha}|^2}{\Delta^2\left(\xi_1,\xi_2,1,\xi_{\phi^+},\xi_\alpha\right)}
\Bigg],
\label{eq:edm_st}
\end{align}
where $\Delta(\xi_i,\xi_j,\xi_\varphi,\xi_{\phi^+},\xi_\alpha)\equiv xy\left(s\xi_i+u\xi_\varphi\right)+xt\xi_{\phi^+}+yt\xi_j-xyu(1-u)\xi_\alpha$ with the dimensionless parameter $\xi_a\equiv m_{a}^2/m_{\eta^+}^2$.
Note that we have taken into account only the rainbow-type diagrams here, 
and some additional diagrams which are topologically different from the rainbow-type ones may give another contribution comparable or even dominant over Eq.~(\ref{eq:edm_st}).

The magnitude of the electron EDM is simply estimated assuming the same order of the fermion masses ($m_e\ll m_1\sim m_2$), the scalar masses ($m_e\ll m_{\eta^+}\sim m_{\phi^+}\sim m_{H}\sim m_{A}$) 
and the Yukawa coupling ($|y_{1e}|\sim |y_{2e}|$). 
In this case, we have numerically checked that the loop functions are approximated by
\begin{align}
&~\int_{0}^{1}\!dxdy\delta(x+y-1)\int_{0}^{1}\!dsdtdu\delta(s+y+u-1)
\frac{x^2y^2su}{\Delta^2\left(\xi_1,\xi_1,1,1,0\right)},\nonumber\\
\approx&\left\{
\begin{array}{ccc}
0.15 & \text{for} & \xi_1\ll 1\\
0.15\xi_1^{-2} & \text{for} & \xi_1\gg 1
\end{array}
\right.,\\
&~\int_{0}^{1}\!dxdy\delta(x+y-1)\int_{0}^{1}\!dsdtdu\delta(s+y+u-1)
\frac{xy^2su}{\Delta^2\left(\xi_1,\xi_1,1,1,0\right)},\nonumber\\
\approx&\left\{
\begin{array}{ccc}
0.25\log\xi_1^{-1} & \text{for} & \xi_1\ll 1\\
0.30\xi_1^{-2} & \text{for} & \xi_1\gg 1
\end{array}
\right.,
\end{align}
therefore the electron EDM is
\begin{align}
\frac{d_e}{e}\approx-\frac{\mathrm{Im}\lambda^2|y_{1e}|^2}{(4\pi)^4}\frac{m_e}{m_{\eta^+}^2}\xi_1\times\left\{
\begin{array}{ccc}
0.25\log\xi_1^{-1} & \text{for} & \xi_1\ll 1\\
0.30\xi_1^{-2} & \text{for} & \xi_1\gg 1
\end{array}
\right..
\end{align}
Fixing the mass ratio $\xi_1=0.1$ and $10$ for examples, the magnitude of the electron EDM is estimated as
\begin{align}
\frac{|d_e|}{e}=\left\{
\begin{array}{ccc}
\displaystyle
1.7\times10^{-30}~\mathrm{cm}\left(\frac{\mathrm{Im}\lambda^2}{0.3}\right)\left(\frac{|y_{1e}|}{0.5}\right)^2
\left(\frac{1~\mathrm{TeV}}{m_{\eta^+}}\right)^2
&\text{for}
&\xi_1=0.1\\
\displaystyle
9.1\times10^{-30}~\mathrm{cm}\left(\frac{\mathrm{Im}\lambda^2}{0.3}\right)\left(\frac{|y_{1e}|}{0.5}\right)^2
\left(\frac{1~\mathrm{TeV}}{m_{1}}\right)^2
&\text{for}
&\xi_1=10
\end{array}
\right..
\end{align}
These values are comparable to the future prospect of the ACME Collaboration that is $ \mathcal{O}(10^{-30})~\mathrm{cm}$~\cite{Kara:2012ay, edm_future}.

\section{The other generalizations}
\label{sec:4}

\begin{figure}[t]
\begin{center}
\includegraphics[scale=1.1]{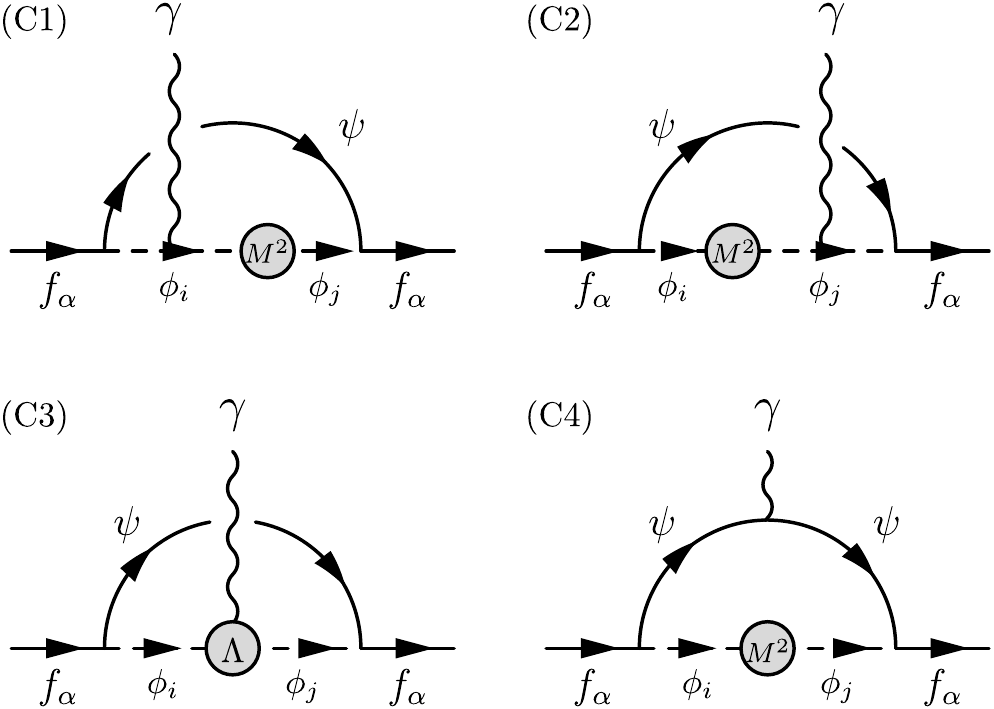}
\caption{Feynman diagrams which may contribute to EDMs of the fermions $f_\alpha$ with multi-scalars $\phi_i$ and a fermion $\psi$.}
\label{fig:diagram_scalar}
\end{center}
\end{figure}

Another generalization of the EDM reduction is possible in models with 
multi-scalars $\phi_i$ and a fermion $\psi$ instead of multi-fermions $\psi_i$ and a scalar $\phi$, or a vector boson $X$.
The Feynman diagrams relevant to the EDM calculation are shown 
in Fig.~\ref{fig:diagram_scalar}.
The calculation for this case is essentially the same as Sec.~\ref{sec:2}. 
We take the effective Lagrangian given by
\begin{align}
 \mathcal{L}_\mathrm{eff}=&-\Big(y_{i\alpha}\phi^*_i\overline{\psi}P_Lf_\alpha+\mathrm{h.c.}\Big)
-\phi_i^*M^2_{ij}\phi_j
-eA_{\mu}\phi_i^*\Lambda_{ij}^{\mu}\phi_j,\nonumber\\
&-eQ_{\psi}A_{\mu}\overline{\psi}\gamma^{\mu}\psi
+ieQ_{\phi}A_{\mu}\big(
\phi_i\partial^{\mu}\raisebox{0.33cm}{\tiny{$\hspace{-0.4cm}\leftrightarrow$}\hspace{0.1cm}}\phi_i^*
\big),
\end{align}
where $M_{ij}^2$ is the self-energy of the scalar particles and $\Lambda_{ij}^{\mu}$ is the vertex correction induced at the loop level
\begin{align}
 \Lambda_{ij}^{\mu}(k_1,k_2)=\sum_{p}Q_p\big(\Lambda_p\big)_{ij}^{\mu}(k_1,k_2).
\end{align}
For the self-energy, the hermiticity condition
\begin{align}
M_{ji}^2(k^2)=M_{ij}^{*2}(k^2),
\label{eq:herm2}
\end{align}
is imposed as same as Sec.~\ref{sec:2}. 

The amplitude for each diagram in Fig.~\ref{fig:diagram_scalar} can be evaluated as
\begin{align}
 i\mathcal{M}_\mathrm{C1}=&~
eQ_\phi\int\!\frac{d^4k}{(2\pi)^4}
 \tilde{M}_{ji}^2(k_2^2)\overline{u}(p_2)P_R\left(\slashed{k}-\slashed{p}+m_\psi\right)P_Lu(p_1)\epsilon_{\mu}\left(k_1+k_2\right)^{\mu}\nonumber\\
 &\times\frac{1}{(k-p)^2-m_\psi^2}\frac{1}{k_1^2-m_i^2}\frac{1}{k_2^2-m_i^2}\frac{1}{k_2^2-m_j^2},\\
i\mathcal{M}_\mathrm{C2}=&~
eQ_\phi\int\!\frac{d^4k}{(2\pi)^4}
 \tilde{M}_{ji}^2(k_1^2)\overline{u}(p_2)P_R\left(\slashed{k}-\slashed{p}+m_\psi\right)P_Lu(p_1)\epsilon_{\mu}\left(k_1+k_2\right)^{\mu}\nonumber\\
 &\times\frac{1}{(k-p)^2-m_\psi^2}\frac{1}{k_1^2-m_i^2}\frac{1}{k_1^2-m_j^2}\frac{1}{k_2^2-m_j^2},\\
i\mathcal{M}_\mathrm{C3}=&~
e\int\!\frac{d^4k}{(2\pi)^4}\epsilon_{\mu}
\tilde{\Lambda}_{ji}^{\mu}(k_1,k_2)
\overline{u}(p_2)P_R\left(\slashed{k}-\slashed{p}+m_\psi\right)P_Lu(p_1)\frac{1}{k_1^2-m_i^2}\frac{1}{k_2^2-m_j^2},\\
i\mathcal{M}_\mathrm{C4}=&~eQ_{\psi}\int\!\frac{d^4k}{(2\pi)^4}
\tilde{M}_{ji}^2(k^2)\overline{u}(p_2)P_R\big(\slashed{p}_2-\slashed{k}+m_\psi\big)\gamma^{\mu}\big(\slashed{p}_1-\slashed{k}+m_\psi\big)P_Lu(p_1)\nonumber\\
&\times \frac{1}{k^2-m_i^2}\frac{1}{k^2-m_j^2}\frac{1}{(p_1-k)^2-m_\psi^2}\frac{1}{(p_2-k)^2-m_\psi^2},
\end{align}
where $\tilde{M}_{ji}^2(k^2)$ and $\tilde{\Lambda}_{ji}^{\mu}(k_1,k_2)$ are defined by
\begin{align}
 \tilde{M}_{ji}^2(k^2)\equiv y_{j\alpha}^* M_{ij}^2(k^2) y_{i\alpha},\qquad
\Lambda_{ji}^{\mu}(k_1,k_2)\equiv y_{j\alpha}^*\Lambda_{ji}^{\mu}(k_1,k_2)y_{i\alpha}.
\end{align}
As in the previous calculation, the scalar vertex correction $\Lambda^{\mu}(k_1,k_2)$ can be decomposed into the longitudinal part and transverse part which can be given by~\cite{Ball:1980ay}
\begin{align}
\Lambda_\mathrm{L}^{\mu}(k_1,k_2)=&~
\frac{M^2(k_1^2)-M^2(k_2^2)}{\left(k\!\cdot\!q\right)}k^{\mu},\\
\Lambda_\mathrm{T}^{\mu}(k_1,k_2)=&~
C(k_1,k_2)\Big[\left(k\!\cdot\!q\right)q^{\mu}-q^2k^{\mu}\Big],
\label{eq:trans_scalar}
\end{align}
where $k_1=k-q/2$ and $k_2=k+q/2$. 
These correspond to Eqs.~(\ref{eq:long}) and (\ref{eq:trans}) in the multi-fermion case.
Note that the vertex correction can be expanded by two vectors $k^{\mu}$ and $q^{\mu}$. 
Then similarly to Sec.~\ref{sec:2}, one vector is removed by the condition of the Ward-Takahashi identity. 
As a result, we obtain the above decomposition. 
One can explicitly find that the longitudinal and transverse parts of the vertex correction obey
\begin{align}
 q_{\mu}\Lambda_\mathrm{L}^{\mu}(k_1,k_2)=&~M^2(k_1^2)-M^2(k_2^2),\\
 q_{\mu}\Lambda_\mathrm{T}^{\mu}(k_1,k_2)=&~0
\end{align}
and thus the Ward-Takahashi identity
\begin{align}
 q_{\mu}\Lambda^{\mu}(k_1,k_2)=&~M^2(k_1^2)-M^2(k_2^2).
\end{align}
From Eq.~(\ref{eq:trans_scalar}), it is clear that the transverse part $\Lambda_\mathrm{T}^{\mu}(k_1,k_2)$ is $\mathcal{O}(q^2)$, and does not contribute to EDMs. 
Thus expanding the full vertex correction $\Lambda^{\mu}(k_1,k_2)$ up to $\mathcal{O}(q)$, one obtains
\begin{align}
 \Lambda^{\mu}(k_1,k_2)\approx \Lambda_\mathrm{L}^{\mu}\Big|_{q=0}+q_{\nu}\frac{\partial\Lambda_\mathrm{T}^{\mu}}{\partial q_{\nu}}\Big|_{q=0}=
-\frac{dM^2}{dk_{\mu}}.
\label{eq:wti_scalar}
\end{align}

The CP violating part of the amplitude arising from (C1) and (C2) diagrams in Fig.~\ref{fig:diagram_scalar} can be evaluated as
\begin{align}
i\mathcal{M}_{\mathrm{C1+C2}}^\mathrm{CP}=&
 -2eQ_{\phi}m_j^2\int\!\frac{d^4k}{(2\pi)^4}\left(k\!\cdot\!q\right)\mathrm{Im}\frac{d\tilde{M}_{ji}^2}{dk^2}
\overline{u}(p_2)P_R\left(\slashed{k}-\slashed{p}+m_{\psi}\right)P_Lu(p_1)
2\left(\epsilon\!\cdot\!k\right)\nonumber\\
&\times
\frac{1}{(k-p)^2-m_\psi^2}\frac{1}{\big(k^2-m_i^2\big)^2}\frac{1}{\left(k^2-m_j^2\right)^2},
\label{eq:amp12cp_scalar}
\end{align}
at $\mathcal{O}(q)$. 
For the amplitude of the diagram (C3), it is exactly the same as Eq.~(\ref{eq:amp12cp_scalar}) except for the opposite sign. 
The amplitude of the diagram (C4) is totally real and does not give CP violation due to the hermiticity condition of Eq.~(\ref{eq:herm2}). 
Therefore summarizing the calculation, it turns out that the CP violating part of the total amplitude in Fig.~\ref{fig:diagram_scalar} exactly vanishes. 

One can also consider multi-vector bosons $X_{i}$ instead of multi-scalars $\phi_i$, and derive the reduced EDM formula. 
However, since there are no well-motivated models involving the multi-vector bosons, we do not consider this case.

\section{Conclusions}
\label{sec:conclusion}

In this paper, we have derived the reduced formulas for the EDM calculation at two-loop level or higher 
introducing the form factors of the self-energy and vertex correction that are related through the Ward-Takahashi identity.
The formulas indicate that one may extract only a part of the self-energy and vertex correction in order to evaluate the fermion EDM arising from the corresponding Feynman diagrams. 
The formulas can be employed in a larger class of models with multi-fermions and a scalar (or a vector boson), and significantly reduces the EDM calculation compared to the usual way. 
We find it is essential to pick up the chirality flipping effects from the internal loop 
in order to obtain the non-zero EDM contributions
as shown in our reduction formulas. 
From this fact, 
we immediately find that EDMs exactly cancel if we consider a model with no chirality flipping effects in the self-energy and the vertex correction for the internal fermion. 

As for an application of our EDM formulas to specific models, first we have revisited the EDM calculation in the SM and the scotogenic model. 
It has been easily verified that the neutron EDM induced by the $W$ bosons vanishes at the two-loop level in the SM. 
Similarly, in the scotogenic model, the charged lepton EDM induced by the relevant rainbow-type diagrams also vanishes. 
In addition, we have considered the singlet-triplet model extended with a triplet scalar to demonstrate how we can obtain a non-zero EDM from our reduction formulas. 

Although the derived EDM formulas can be adapted in a large class of models, one should note that additional diagrams of different topology which are not treated in this paper may exist. 
For example, in the scotogenic model~\cite{Fujiwara:2020unw}, different topological diagrams exist and give  a leading contribution to the EDM.

\section*{Acknowledgments}
\noindent
MF would like to thank Teppei Kitahara for the useful discussion. 
This work was supported by JSPS Grant-in-Aid for Scientific Research KAKENHI Grant No. JP20J12392 (MF), JP20H01895 (JH), JP21K03572 (JH), and JP20K22349 (TT).
The work of J.H. was supported by Grant-in-Aid for Scientific research from the Ministry of Education, Science, Sports, and Culture (MEXT), Japan (Grant Numbers 16H06492).
The work of J.H. was also supported by JSPS Core-to-Core Program (Grant Numbers JPJSCCA20200002),
and World Premier International Research Center Initiative (WPI Initiative), MEXT, Japan.

\appendix
\section{Decomposition of vertex correction}
\label{sec:append1}
One can take the following 12 independent vectors to construct the vertex correction $\Lambda^{\mu}(k_1,k_2)$~\cite{Bernstein:1968aa},
\begin{alignat}{4}
\mathcal{V}_1^{\mu}=&~k^{\mu},\quad\hspace{0.67cm}
\mathcal{V}_2^{\mu}=q^{\mu},\quad\hspace{2.22cm}
\mathcal{V}_3^{\mu}=\gamma^{\mu},\quad\hspace{0.5cm}
\mathcal{V}_4^{\mu}=i\sigma^{\mu\nu}k_{\nu},\nonumber\\
\mathcal{V}_5^{\mu}=&~i\sigma^{\mu\nu}q_{\nu},\quad
\mathcal{V}_6^{\mu}=\slashed{k}q^{\mu},\quad\hspace{1.98cm}
\mathcal{V}_7^{\mu}=\slashed{q}k^{\mu},\quad\hspace{0.3cm}
\mathcal{V}_8^{\mu}=\slashed{k}k^{\mu},\nonumber\\
\mathcal{V}_9^{\mu}=&~\slashed{q}q^{\mu},\quad\hspace{0.41cm}
\mathcal{V}_{10}^{\mu}=i\epsilon^{\mu\nu\rho\sigma}\gamma_5\gamma_{\nu}k_{\rho}q_{\sigma},\quad
\mathcal{V}_{11}^{\mu}=\slashed{k}\slashed{q}k^{\mu},\quad
\mathcal{V}_{12}^{\mu}=\slashed{k}\slashed{q}q^{\mu},
\label{eq:append1}
\end{alignat}
where $q=k_2-k_1$, $k_1=k-q/2$ and $k_2=k+q/2$. 
Then the vertex correction $\Lambda^{\mu}(k_1,k_2)$ is given by
\begin{align}
\Lambda^{\mu}(k_1,k_2)=\sum_{a=1}^{12}\Big[\mathcal{C}_{a}^{L}(k_1,k_2)\mathcal{V}_{a}^{\mu}P_L+\mathcal{C}_{a}^{R}(k_1,k_2)\mathcal{V}_{a}^{\mu}P_R\Big].
\label{eq:append2}
\end{align}
Multiplying the photon momentum $q_{\mu}$ to Eq.~(\ref{eq:append2}), it should satisfy the Ward-Takahashi identity shown in Eq.~(\ref{eq:wti}). 
Using the explicit form of the self-energy in Eq.~(\ref{eq:self-energy}), and comparing the coefficients of the terms proportional to $\slashed{k}\slashed{q}$, $\slashed{k}$, $\slashed{q}$ and $1$, one can obtain the following 4 equations for the left chirality
\begin{align*}
\mathcal{C}_4^{L}+\mathcal{C}_{11}^{L}\left(k\!\cdot\!q\right)+\mathcal{C}_{12}^{L}q^2=&~0,\\
\mathcal{C}_6^{L}q^2+\mathcal{C}_8^{L}\left(k\!\cdot\!q\right)=&~A^{L}(k_1^2)-A^{L}(k_2^2),\\
\mathcal{C}_3^{L}+\mathcal{C}_7^{L}\left(k\!\cdot\!q\right)+\mathcal{C}_9^{L}q^2=&-\frac{A^{L}(k_1^2)+A^{L}(k_2^2)}{2},\\
\mathcal{C}_1^{L}\left(k\!\cdot\!q\right)+\mathcal{C}_2^{L}q^2-\mathcal{C}_4^{L}\left(k\!\cdot\!q\right)=&~B^{L}(k_1^2)-B^{L}(k_2^2),
\end{align*}
and the same equations for the other chirality are also obtained. 
Note that for the vectors 
$\mathcal{V}_5^{\mu}$ and $\mathcal{V}_{10}^{\mu}$, it gives identically zero when $q_{\mu}$ is multiplied. 
From these equations, 4 coefficients for each chirality can be eliminated, which are chosen as $\mathcal{C}_1^{L/R}$, $\mathcal{C}_3^{L/R}$, $\mathcal{C}_8^{L/R}$ and $\mathcal{C}_{11}^{L/R}$ here. 
As a result, one can decompose the vertex correction into the longitudinal part 
\begin{align}
 \Lambda_\mathrm{L}^{\mu}(k_1,k_2)=&~
\Bigg[\frac{A^L(k_1^2)-A^L(k_2^2)}{\left(k\!\cdot\!q\right)}k^{\mu}\slashed{k}
-\frac{A^L(k_1^2)+A^L(k_2^2)}{2}\gamma^{\mu}
+\frac{B^L(k_1^2)-B^L(k_2^2)}{\left(k\!\cdot\!q\right)}k^{\mu}\Bigg]P_L\nonumber\\
&~+\left(L\leftrightarrow R\right),
\end{align}
which corresponds to Eq.~(\ref{eq:long}), and the transverse part 
\begin{align}
\Lambda_\mathrm{T}^{\mu}(k_1,k_2)=&~
\frac{\mathcal{C}_2^{L}}{\left(k\!\cdot\!q\right)}\big[\left(k\!\cdot\!q\right)q^{\mu}-q^2k^{\mu}\big]P_L
+\frac{\mathcal{C}_4^{L}}{\left(k\!\cdot\!q\right)}\slashed{k}\big[\left(k\!\cdot\!q\right)\gamma^{\mu}-\slashed{q}k^{\mu}\big]P_L
+\mathcal{C}_5^{L}i\sigma^{\mu\nu}q_{\nu}P_L\nonumber\\
&~+\frac{\mathcal{C}_6^{L}}{\left(k\!\cdot\!q\right)}\slashed{k}\big[\left(k\!\cdot\!q\right)q^{\mu}-q^2k^{\mu}\big]P_L
-\mathcal{C}_7^{L}\big[\left(k\!\cdot\!q\right)\gamma^{\mu}-\slashed{q}k^{\mu}\big]P_L
+\mathcal{C}_9^{L}\big[\slashed{q}q^{\mu}-q^2\gamma^{\mu}\big]P_L\nonumber\\
&~+\mathcal{C}_{10}^{L}\epsilon^{\mu\nu\rho\sigma}\gamma_5\gamma_{\nu}k_{\rho}q_{\sigma}P_L
+\frac{\mathcal{C}_{12}^{L}}{\left(k\!\cdot\!q\right)}\slashed{k}\slashed{q}\big[\left(k\!\cdot\!q\right)q^{\mu}-q^2k^{\mu}\big]P_L
+\left(L\rightarrow R\right).
\label{eq:append3}
\end{align}
Finally, replacing the coefficients
\begin{align}
\frac{\mathcal{C}_2^{L/R}}{\left(k\!\cdot\!q\right)}&\to C_1^{L/R},\quad\hspace{0.4cm}
\frac{\mathcal{C}_4^{L/R}}{\left(k\!\cdot\!q\right)}\to C_6^{L/R},\quad
\mathcal{C}_5^{L/R}\to C_7^{L/R},\quad
\frac{\mathcal{C}_6^{L/R}}{\left(k\!\cdot\!q\right)}\to C_2^{L/R},\nonumber\\
-\mathcal{C}_7^{L/R}&\to C_5^{L/R},\quad
~~~~\mathcal{C}_9^{L/R}\to C_4^{L/R},\quad
\mathcal{C}_{10}^{L/R}\to C_8^{L/R},\quad
\frac{\mathcal{C}_{12}^{L/R}}{\left(k\!\cdot\!q\right)}\to C_3^{L/R},
\end{align}
Eq.~(\ref{eq:append3}) eventuates in Eq.~(\ref{eq:trans}).


\end{document}